\documentstyle[psfig]{l-aa}
                    
\begin{document}

\thesaurus{
       (12.03.3;  
        12.12.1;  
Universe
        11.03.1;  
         ) }


\title{The ESO Nearby Abell Cluster Survey
       \thanks{Based on observations collected at the 
European Southern
               Observatory (La Silla, Chile)}
       \thanks{http://www.astrsp-mrs.fr/www/enacs.html}
      }

\subtitle{VII: Galaxy density profiles of rich clusters of galaxies}

\author{C.~Adami \inst{1}, A.~Mazure \inst{1} , P. Katgert \inst{2}, 
A. Biviano \inst{3}}

\institute{IGRAP, Laboratoire d'Astronomie Spatiale, 
Marseille, France 
   \and Sterrewacht Leiden, The Netherlands 
   \and Osservatorio Astronomico di Trieste, Italy }

\offprints{C.~Adami}
\date{Received date; accepted date}

\maketitle

\markboth{VII: Density profiles of rich clusters of galaxies}{}

\begin{abstract}

We have analyzed the projected galaxy distributions in a subset of the
ENACS cluster sample, viz. in those 77 clusters that have z $<$ 0.1
and R$_{\rm ACO} \ge$ 1 and for which ENACS and COSMOS data are
available.  For 20~\% of these, the distribution of galaxies in the
COSMOS catalogue does not allow a reliable centre position to be
determined.  For the other 62 clusters, we first determined the centre
and elongation of the galaxy distribution. Subsequently, we made
Maximum-Likelihood fits to the distribution of COSMOS galaxies for 4
theoretical profiles, two with `cores' (generalized King- and
Hubble-profiles) and two with `cusps' (generalized Navarro, Frenk and
White, or NFW, and de~Vaucouleurs profiles).

We obtain average core radii (or characteristic radii for the 
profiles without core) of 128, 189, 292 and 1582 kpc for fits
with King, Hubble, NFW and de~Vaucouleurs profiles respectively, with
dispersions around these average values of 88, 116, 191 and 771 kpc.
The surface density of background galaxies is about 4 10$^{-5}$ gals
arcsec$^{-2}$ (with a spread of about 2 10$^{-5}$), and there is very
good agreement between the values found for the 4 profiles. There is
also very good agreement on the outer logarithmic slope of the
projected galaxy distribution, which is that for the non-generalized
King- and Hubble-profile (i.e.  $\beta_{King}$ = $\beta_{Hubble}$ = 1,
with the corresponding values for the two other model-profiles).

We use the Likelihood ratio to investigate whether the observations
are significantly better described by profiles with cusps or by
profiles with cores. Taking the King and NFW profiles as `model' of
either class, we find that about 75 \% of the clusters are better fit
by the King profile than by the NFW profile.  However, for the
individual clusters the preference for the King profile is rarely
significant at a confidence level of more than 90 \%.  When we limit
ourselves to the central regions it appears that the signifance
increases drastically, with 65 \% of the clusters showing a strong
preference for a King over an NFW profile. At the same time, about 10
\% of the clusters are clearly better fitted by an NFW profile than by a
King profile in their centres.

We constructed composite clusters from the COSMOS and ENACS data,
taking special care to avoid the creation of artificial cusps (due to
ellipticity), and the destruction of real cusps (due to non-perfect
centering). When adding the galaxy distributions to produce a
composite cluster, we either applied no scaling of the projected
distances, scaling with the core radii of the individual clusters or
scaling with r$_{200}$, which is designed to take differences in mass
into account. In all three cases we find that the King profile is
clearly preferred (at more than 95 \% confidence) over the NFW profile
(over the entire aperture of 5 core-radii). However, this `preference'
is not shared by the brightest (M$_{b_j}$ $\la$ -18.4) galaxies. We
conclude that the brighter galaxies are represented almost equally
well by King and NFW profiles, but that the distribution of the
fainter galaxies clearly shows a core rather than a cusp.
  
Finally, we compared the outer slope of the galaxy distributions in
our clusters with results for model calculations for various choices
of fluctuation spectrum and cosmological parameters. We conclude that
the observed profile slope indicates a low value for $\Omega_0$. This
is consistent with the direct estimate of $\Omega_0$ based on the
$\frac ML$-ratios of the individual clusters.

\end{abstract}


\section{Introduction}

Until fairly recently, the projected galaxy density in rich galaxy
clusters was generally described by King or Hubble profiles. In these
profiles, the logarithmic slope of the mass distribution is
essentially zero near the cluster centre. The core radius which is the
characteristic scale of the distribution, was sometimes also regarded
as the distance which more or less separates dynamically distinct
regions in a cluster. From the kinematics of the galaxy population it
appears that in clusters the relaxation time is significantly shorter
than the Hubble time {\em only} in the very central region within at
most a few core radii (see e.g. den Hartog and Katgert 1996).

The concept of cores in clusters has been seriously challenged, on
observational grounds (e.g. Beers $\&$ Tonry 1986) and as a result of
numerical simulations. Navarro, Frenk and White (1995, 1996) found
e.g. that the equilibrium density profiles of dark matter halos in
universes with dominant hierarchical clustering all have the same
shape, which is essentially independent of the mass of the halo, the
spectrum of initial density fluctuations, or the values of the
cosmological parameters. This `universal' density profile (NFW profile
hereafter) does not have a core, but has a logarithmic slope of --1
near the centre which, at large radii, steepens to --3, and thus
closely resembles the Hernquist (1990) profile except for the steeper
slope of the latter at large radii of --4.

Navarro, Frenk and White (1997) argue that the apparent variations in
profile shape, as reported before, can be understood as being due to
differences in the characteristic density (or mass) of the halo, which
sets the linear scale at which the transition of the flat central
slope to the steep outer slope occurs. They also argued that the
existence of giant arcs in clusters requires that the mass
distributions in clusters does not exhibit a flat core in the
centre. In other words: if clusters have cores, the lensing results
require that the core radii are very small, at least quite a bit
smaller than the values usually quoted.

It is not clear that galaxy clusters should have cores; after all, the
dynamical structure of galaxy clusters is quite different from that of
globular clusters, for which Michie \& Bodenheimer (1963) and King
first proposed density profiles with cores, in particular the King
profile (see e.g. King 1962). On the other hand, the X-ray data for
clusters are quite consistent with the existence of a core in the
density distribution. More specifically, it was argued recently by
Hughes (1997) that the NFW profile would induce a temperature
gradient. The existence of such a gradient in the Coma cluster can be 
excluded at the 99$\%$ confidence level. 
Similarly, the galaxy surface density in
clusters is generally found to be consistent with a King profile. For
galaxy clusters, little use has been made of the de Vaucouleurs
profile to describe the galaxy density, even though the latter was
found to arise quite naturally in N-body simulations of the collapse
of isolated galaxy systems (e.g. van Albada 1982).

In view of the claimed universality of the NFW profile found in the
simulations, it seems useful to have a closer look at the projected
distribution of the galaxies in clusters. After all, the NFW profile
refers to the total gravitating mass, and it is not obvious that the
galaxy distribution should have exactly the same shape as the
distribution of total mass; although in numerical experiments no
strong biasing between dark and luminous matter in clusters was seen
(e.g. van Kampen 1995). In this respect, it is noteworthy that
Carlberg et al. (1997) find that the combined galaxy density profile
of 16 high-luminosity X-ray clusters at a redshift of $\approx$0.3
closely follows the NFW profile. More precisely, the logarithmic slope
in the central region is consistent with the value of --1 of the NFW
and Hernquist profiles, while the outer slope is consistent with both
--3 (the NFW value) and --4 (the Hernquist value).

The outer slope of the density profile was found by several authors
(e.g. Crone et al. 1994, Jing et al. 1995, and Walter and Klypin 1996)
to reflect the details of the formation scenario, and in particular
the value of the density parameter of the universe. In addition, this
slope is unlikely to be constant in time but is expected to get
steeper with decreasing redshift. For that reason, it is important to
study both the characteristics of the density profiles of rich
clusters and their dependence on redshift.

In this paper, we investigate the projected galaxy distributions for a
sample of 62 rich and nearby (z $\la$ 0.1) clusters. These
clusters are taken from the volume-limited ENACS (ESO Nearby Abell
Cluster Survey) sample of R$_{\rm ACO} \ge 1$ clusters (see e.g.
Katgert et al. 1996 (paper I), Mazure et al.  1996 (paper II), Biviano
et al. 1997 (paper III), Adami et al. 1998 (paper IV), Katgert et
al. 1998 (paper V) and de Theije $\&$ Katgert 1998 (paper VI)). In $\S$ 2, 
we first describe the sample of
clusters that we used, and the data on which we based our analysis. In
$\S$ 3, we discuss the results of Maximum-Likelihood fitting of
profiles with and without a core, to the individual galaxy
distributions taken from the COSMOS catalogue. In $\S$ 4 we discuss
the galaxy density distribution for composite clusters (COSMOS and ENACS),
 in which the individual clusters are combined.  In $\S$ 5 we discuss the
constraints provided by the outer slope of the density distributions
for the parameters of the formation scenario and in $\S$ 6 we present
the conclusions.

\section{The data}

\subsection{The galaxy catalogues}

In this paper we use both COSMOS and ENACS data. The COSMOS data, i.e.
photometric galaxy catalogues that were obtained from automatic
scanning of UK Schmidt IIIa--J survey plates with the Edinburgh
plate-scanning machine, were kindly provided by Dr. H.T.
MacGillivray.  As described in paper V, the COSMOS data that we used
are of two kinds: the well-calibrated part around the Southern
Galactic Pole (the so-called EDSGC, see e.g.  Heydon-Dumbleton et
al. 1989), and the slightly less well-calibrated remainder. In paper
V, we compared the COSMOS and ENACS photometry and found only a small
difference in the calibration quality of the two COSMOS subsets. We
did not find evidence for systematic magnitude offsets between the two
COSMOS subsets, nor for differences in completeness. From a comparison
with the ENACS galaxy catalogues, which are based on completely
independent scanning with the Leiden Observatory plate-measuring
machine, we concluded that the COSMOS catalogue is 90\% complete to a
nominal limit of m$_{b_j}$ $\approx$ 19.5.

The galaxy catalogues that resulted from the ENACS spectroscopic
survey are described in papers I and V. Very briefly, redshifts were
obtained for 5634 galaxies in 107 ACO cluster candidates, mostly in
the southern hemisphere. As shown in paper II, the ENACS allowed us to
construct a complete, volume-limited sample of 128 rich (R$_{\rm ACO}
\ge$ 1) clusters, out to a redshift of 0.1, when we combine ENACS data
for about 80 clusters with literature data for about 50 clusters. In
paper V, we have shown the magnitude distributions of the galaxy
samples for which we obtained ENACS spectroscopy.  Comparison with the
magnitude distributions of the COSMOS galaxies shows the ENACS samples
to be more or less complete to m$_{b_j}$ between about 18 and 19, with
quite a few redshifts for fainter galaxies, viz. down to m$_{b_j}
\approx$ 19.5.

\subsection{The cluster sample}

The overlap of the ENACS dataset with that part of the COSMOS dataset
that was available to us yields a sample of 77 clusters.  These
clusters were not selected according to particular criteria, and we
therefore expect them to be a representative subset of the total ENACS
sample. To this sample, we have added the cluster A2721, for which we
use redshift data from Colless $\&$ Hewett (1987), Colless (1989) and
Teague et al. (1990).

Because we want to study density profiles, we are reluctant to use
clusters with clear signs of substructure, as for the latter a scale
length and the inner and outer slopes of the density profile do not
have well-defined meanings. In addition, it is not at all trivial to
define a meaningful centre for systems that appear irregular in
projection.

In order to give a visual overview of the galaxy distributions in the
77 clusters in the sample, we have used the COSMOS data to produce
adaptive-kernel maps for all 77 clusters; these are shown in Figs. 1
and 2.

\begin{figure*}
\vbox
{\psfig{file=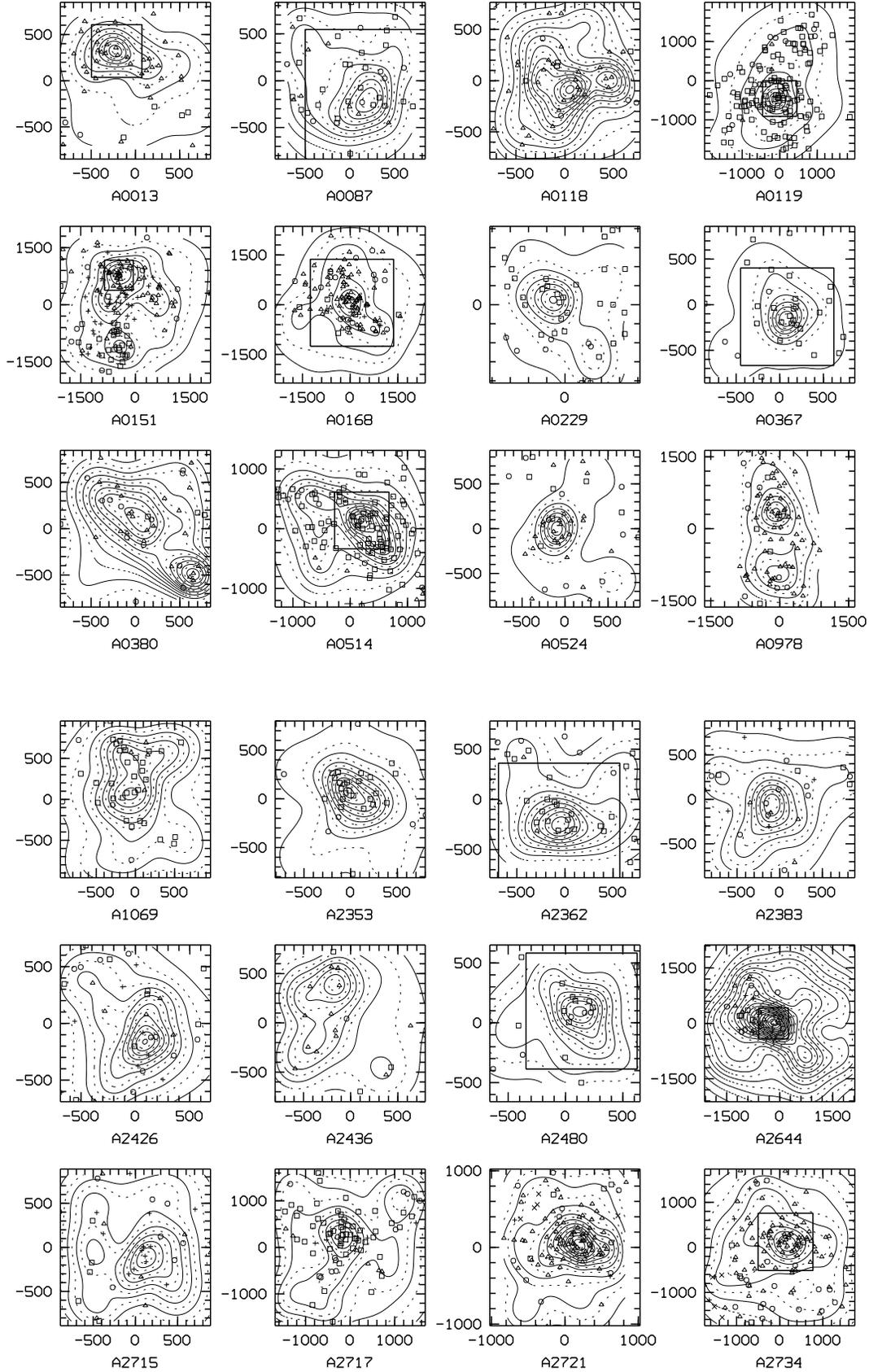,width=16.5cm,height=23.5cm,angle=90}}
\caption[]{Adaptive-kernel maps of projected galaxy density from 
the COSMOS catalogue. Coordinates are in arcsec with respect to the
cluster centre. Different groups of redshift are indicated as follows:
nearest group on the line of sight: circles, 2nd group: squares, 3rd:
triangles, 4th: crosses, 5th: asterisks and 6th: stars. 
We have indicated the area we use for the 29 clusters involved in the COSMOS 
composite clusters. For A1069, we use the whole area.}
\end{figure*}

\begin{figure*}
\addtocounter{figure}{-1}
\vbox
{\psfig{file=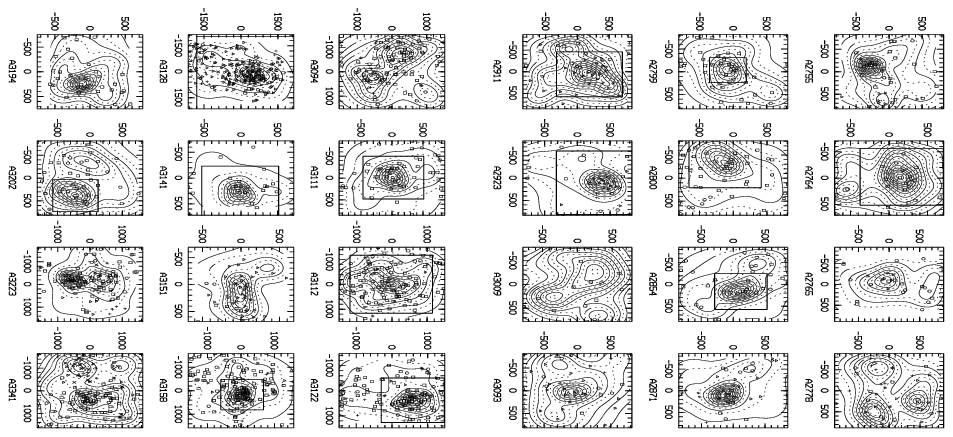,width=16.5cm,height=23.5cm,angle=90}}
\caption[]{continued}
\end{figure*}

\begin{figure*}
\addtocounter{figure}{-1}
\vbox
{\psfig{file=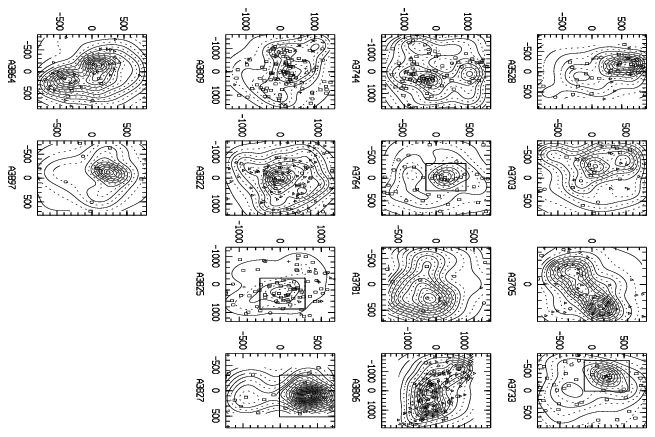,width=16.5cm,height=17.0cm,angle=90}}
\caption[]{continued}
\end{figure*}

\begin{figure*}
\vbox
{\psfig{file=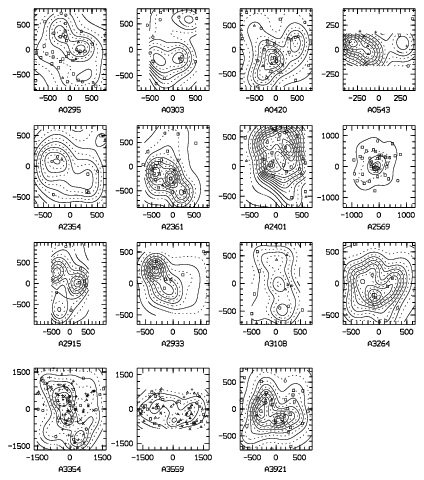,width=18.0cm,height=17cm,angle=90}}
\caption[]{Adaptive-kernel maps of projected galaxy density from the 
COSMOS catalogue for the 15 clusters for which the centre could not be
determined reliably. Coordinates are in arcsec with respect to the
cluster centre. The different groups of redshift are characterized by
a different symbol: nearest group on the line of sight: circles, 2nd
group: squares, 3rd: triangles, 4th: crosses, 5th: asterisks and 6th:
stars.}
\end{figure*}

The contour maps in these figures are based on surface densities
calculated as the sum of normalized gaussian kernel functions at the
positions of the galaxies with dispersions $\sigma_i$ that are adapted
to the local galaxy density (e.g. Silverman 1986). In Figs. 1 and 2 ,
it is fairly easy to distinguish clusters with single density peaks,
clusters with multiple density peaks that can be clearly separated
and/or identified with systems at different redshifts, and irregular
clusters. By using different symbols for systems with different
redshifts (see papers I and IV), it is possible to recognize clusters
which consist of several `single' peaks at different redshifts. About
two-thirds of the clusters in our sample show a single peak, while
about 10\% was considered irregular. We have defined the studied area
for each cluster as the maximal area without a clear sign of substructure 
and with a single density peak. Afterwards, we have checked that these 
areas comprise more than 5 King core radii (i.e. with a radius greater than 
500 kpc).

In Tab. 1, we give our verdict on the character of the galaxy
distribution, as well as some other information, for the 62 clusters
in Fig. 1 for which a central position could be determined reliably
(see $\S$ 3). The 15 clusters for which no reliable position could be
obtained are shown in Fig. 2; those could not be used in the following
analysis. It should be noted that Tab. 1 has 2 entries for A0151,
while the cluster A3559 (in the region of the Shapley concentration)
was discarded because it did not show a clear maximum in the galaxy
distribution. Note also that our verdict `regular' in Tab. 1 does not
imply that the cluster does not have substructure, as projection may
hide substructure along the line-of-sight.

The mean redshift of the systems in Tab. 1 is about 0.075 (practically
equal to that of the total ENACS sample). In the following analysis
we will adopt $H_0$ = 100 km/s/Mpc and $q_0$ = 0. The parameters 
will be determined with robust statistical
estimators (Beers et al. 1990).

\begin{table} 
\caption[]{Parameters of the 62 clusters for which a reliable centre 
position was obtained. Col.(1) cluster name, col.(2) ENACS redshift of
the group, col (3) type of galaxy distribution (0: regular single
peak, 1: multimodal, 2: irregular), cols.(4) and (5) centre position,
col (6) ellipticity (when available) and col (7) direction of the
major axis (if ellipticity available and not equal to 0), counted
anti-clockwise from $\delta$=0.}
\begin{flushleft}
{\scriptsize 
\begin{tabular}{cccrrcr} 
\hline 
\noalign{\smallskip} 
ACO & z & type & \multicolumn{1}{c}{$\alpha $} &
                 \multicolumn{1}{c}{$\delta $} & e & $ \phi $ \\
 & &           & \multicolumn{2}{c}{2000.0} & & \\ 
\noalign{\smallskip}
\hline
\noalign{\smallskip}
0013 & 0.094 & 0 & 00:13:34.53 & -19:29:38  & 0.32 & -16 
\\ 
0087 & 0.055 & 0 & 00:43:00.73 & -09:50:37  & 0.00 &  \\ 
0118 & 0.115 & 1 & 00:55:21.47 & -26:23:24  & 0.00 &  \\ 
0119 & 0.044 & 0 & 00:56:20.13 & -01:15:51  & 0.00 &  \\ 
0151 & 0.041 & 0 & 01:08:50.07 & -15:25:05  & 0.14 & -25 
\\ 
0151 & 0.053 & 2 & 01:08:52.73 & -15:56:51  & 0.00 &  \\ 
0168 & 0.045 & 0 & 01:15:14.87 &  00:15:23  & 0.00 &  \\ 
0229 & 0.113 & 0 & 00:14:34.47 &  00:01:21  & 0.50 & -45 
\\ 
0367 & 0.091 & 0 & 02:36:35.20 & -19:22:16  & 0.05 &  70 
\\ 
0380 & 0.134 & 1 & 02:44:23.33 & -26:13:58  & 0.12 & -43 
\\ 
0514 & 0.072 & 1 & 04:48:15.13 & -20:27:26  & 0.28 & -25 
\\ 
0524 & 0.078 & 0 & 04:57:47.87 & -19:43:11  &      &  \\ 
0978 & 0.054 & 1 & 10:20:24.33 & -06:29:53  & 0.04 &   0 
\\ 
1069 & 0.065 & 0 & 10:39:47.93 & -08:40:46  & 0.00 &  \\ 
2353 & 0.121 & 0 & 21:34:27.33 & -01:36:15  &      &  \\ 
2362 & 0.061 & 0 & 21:39:03.33 & -14:21:10  & 0.23 & -2 
\\ 
2383 & 0.058 & 0 & 21:52:10.00 & -21:09:35  &      &  \\ 
2426 & 0.098 & 0 & 22:14:35.33 & -10:21:51  & 0.00 &  \\ 
2436 & 0.091 & 2 & 22:20:31.33 & -02:46:57  & 0.21 &  28 
\\ 
2480 & 0.072 & 0 & 22:46:10.67 & -17:41:22  & 0.08 & -52 
\\ 
2644 & 0.069 & 0 & 23:41:02.00 &  00:05:30  & 0.29 & -20 
\\ 
2715 & 0.114 & 0 & 00:02:48.60 & -34:40:57  & 0.00 &  \\ 
2717 & 0.049 & 2 & 00:03:07.73 & -35:56:50  & 0.00 &  \\ 
2721 & 0.120 & 0 & 00:06:12.53 & -34:43:12  & 0.00 &  \\ 
2734 & 0.062 & 0 & 00:11:22.93 & -28:50:55  & 0.00 &  \\ 
2755 & 0.095 & 0 & 00:17:39.20 & -35:11:57  & 0.07 &  27 
\\ 
2764 & 0.071 & 0 & 00:20:29.53 & -49:14:14  & 0.00 &  \\ 
2765 & 0.080 & 0 & 00:21:31.53 & -20:45:55  & 0.00 &  \\ 
2778 & 0.102 & 1 & 00:29:08.67 & -30:17:28  & 0.00 &  \\ 
2799 & 0.063 & 0 & 00:37:24.13 & -39:09:01  & 0.17 &   5 
\\ 
2800 & 0.064 & 0 & 00:37:58.67 & -25:05:17  & 0.37 &  35 
\\ 
2854 & 0.061 & 0 & 01:00:47.20 & -50:32:38  & 0.54 & -68 
\\ 
2871 & 0.122 & 0 & 01:08:08.07 & -36:45:30  & 0.00 &  \\ 
2911 & 0.064 & 0 & 01:26:08.00 & -37:57:26  & 0.00 &  \\ 
2923 & 0.061 & 0 & 01:32:28.80 & -31:04:41  & 0.00 &  \\ 
3009 & 0.120 & 2 & 02:21:33.53 & -48:28:43  &      &  \\ 
3093 & 0.081 & 0 & 03:10:54.93 & -47:23:53  &      &  \\ 
3094 & 0.071 & 1 & 03:12:36.67 & -27:08:05  & 0.00 &  \\ 
3094 & 0.065 & 1 & 03:11:23.13 & -26:54:21  & 0.00 &  \\ 
3111 & 0.083 & 0 & 03:17:49.00 & -45:43:38  & 0.02 &  30 
\\ 
3112 & 0.075 & 0 & 03:17:58.53 & -44:14:21  & 0.30 &  73 
\\ 
3122 & 0.068 & 0 & 03:22:14.00 & -41:19:14  & 0.00 &  \\ 
3128 & 0.078 & 1 & 03:30:37.87 & -52:31:51  & 0.00 &  \\ 
3141 & 0.075 & 0 & 03:36:54.27 & -28:04:20  & 0.44 &  50 
\\ 
3151 & 0.064 & 0 & 03:40:07.80 & -28:41:27  &      &  \\ 
3158 & 0.060 & 0 & 03:43:04.60 & -53:38:40  & 0.00 &  \\ 
3194 & 0.105 & 0 & 03:59:07.20 & -30:11:24  &      &  \\ 
3202 & 0.059 & 0 & 04:00:55.53 & -53:41:17  & 0.13 &  49 
\\ 
3223 & 0.060 & 1 & 04:08:05.80 & -31:03:20  & 0.27 &  61 
\\ 
3341 & 0.038 & 2 & 05:25:34.20 & -31:36:35  & 0.06 &  41 
\\ 
3528 & 0.054 & 0 & 12:54:24.07 & -29:02:05  &      &  \\ 
3703 & 0.074 & 2 & 20:39:52.67 & -61:19:34  &      &  \\ 
3705 & 0.090 & 2 & 20:42:04.00 & -35:13:07  & 0.65 &  35 
\\ 
3733 & 0.039 & 0 & 21:01:34.67 & -28:02:42  & 0.49 &  35 
\\ 
3744 & 0.038 & 1 & 21:07:26.00 & -25:24:36  & 0.00 &  \\ 
3744 & 0.038 & 1 & 21:07:09.33 & -25:01:56  &  &  \\ 
3764 & 0.075 & 0 & 21:25:47.33 & -34:42:44  & 0.35 & -28 
\\ 
3781 & 0.057 & 0 & 21:35:27.33 & -66:49:14  &      &  \\ 
3806 & 0.076 & 2 & 21:46:24.00 & -57:17:28  & 0.22 & -15 
\\ 
3806 & 0.076 & 2 & 21:48:09.33 & -57:17:15  &      &  \\ 
3809 & 0.062 & 2 & 21:47:17.33 & -43:52:54  & 0.00 &  \\ 
3822 & 0.076 & 1 & 21:54:03.33 & -57:50:27  & 0.00 &  \\ 
3825 & 0.075 & 0 & 21:58:26.00 & -60:22:03  & 0.00 &  \\ 
3827 & 0.098 & 0 & 22:01:52.00 & -59:56:42  & 0.00 &  \\ 
3864 & 0.102 & 1 & 22:19:47.33 & -52:27:05  & 0.00 &  \\ 
3897 & 0.073 & 0 & 22:39:10.67 & -17:20:14  & 0.00 &  \\
\hline	   
\end{tabular}
}
\normalsize
\end{flushleft}
\end{table}

\section{The density profiles of the individual clusters}
\subsection{The Maximum-Likelihood fits}

We derived the characteristics of the projected galaxy distributions
for the individual clusters using Maximum-Likelihood fits (hereafter
MLM fits, see e.g. Sarazin 1980 for a description of the method). We
have made MLM fits for different types of profile as follows. Define
$\sigma (x,y)$ as the theoretical, normalized density profile with
which one wants to compare the data. 
Note that we scaled the amplitude of the model to reproduce the observed 
number of galaxies. 

The probability that this assumed
profile `produces' a galaxy in position (x$_k$,y$_k$) is
thus $\sigma (x_k,y_k)$. Consequently, the combined probability that
the assumed profile will `produce' galaxies in the positions
(x$_k$,y$_k$) (with k=1...N) that they actually have is:

L=$\prod\limits_{k=1}^N\sigma (x_k,y_k)$

The model-parameters which produce the best fit between the data and
the model are found from a maximization of the likelihood L. The
combined probability, L, for all galaxies to have their actual
positions assuming the model to be correct, is always less than
unity. For that reason, in practice, one usually minimizes the
positive parameter --2ln(L) w.r.t. the parameters of the assumed
profile.  This yields the values of the parameters for which the model
is most likely to have generated the N galaxies at their observed
positions.  The great advantage of this method is that it does not
require the data to be binned. In addition, it uses all information
that is available.

The minimization of --2ln(L) was done with the two minimization
methods in the MINUIT package: SIMPLEX (Nelder $\&$ Mead, 1965) and
MIGRAD (Fletcher 1970). We use the same stategy as described in paper
IV to obtain convergence. Viz., we use SIMPLEX to approach the optimal
values and MIGRAD to refine those and get reliable error estimates.
When MIGRAD did not converge, we adopted the SIMPLEX value without
error estimates.

We used the following model profiles for the 2D galaxy 
distributions:
 
\noindent - generalized King model: $\sigma (r)=\sigma _0(\frac 1{1+ (
\frac{r}{r_c})^2})^\beta +\sigma_b$,

\noindent - generalized Hubble model: $\sigma (r)=\sigma_0( \frac 1{1+
\frac{r}{r_c}}) ^{2\beta }+\sigma_b$,

\noindent - generalized NFW model: $\sigma (r)=\sigma_0( \frac 1
{\frac{r}{r_c}(1+\frac{r}{r_c}) ^2}) ^{\beta }+\sigma_b$,

Note that the 2-D model profile that we refer to as NFW is not an
exact 2-D version of the 3-D profile described by NFW. Instead, it is
a pseudo NFW profile, with a relation between 3-D and 2-D outer
logartithmic slopes as for the King and Hubble profiles, but with an
inner logarithmic slope equal to that of the 3-D NFW profile.

\noindent - generalized de Vaucouleurs model: $\sigma (r)=\sigma _0
e^{-\beta (\frac{r}{r_c}) ^{0.25}} +\sigma_b$.  \smallskip

\noindent As the actual clusters are not necessarily axi-symmetric,
the models have 7 free parameters: two parameters for the position of
the centre (x$_0$, y$_0$), two parameters to describe deviations from
symmetry (ellipticity e and position angle $\phi$), two parameters
that specify the profile (r$_c$ and $\beta$) and, finally, the
background density $\sigma_b$ (assumed constant within the aperture of
each cluster).

In principle, an MLM solution could have been made for all 7
parameters simultaneously. However, we have separated the solution for
the central position and the elongation and position angle, from the
solution of the background density and the profile parameters r$_c$
and $\beta$. More precisely: we used a King profile with r$_c$ = 100
kpc, $\beta$ = 1.0, and we assumed a value for $\sigma_b$ of 3
10$^{-5}$ galaxies per square arcsec to make an MLM fit for x$_0$,
y$_0$, e, and $\phi$. As mentioned previously, this fit did not
converge for 15 clusters, viz. for A0295, A0303, A0420, A0543, A2354,
A2361, A2401, A2569, A2915, A2933, A3108, A3264, A3354, A3559 and
A3921. These clusters could therefore not be used for MLM fits for the
four profiles.

The average uncertainty in the central position as given by the fits
is about 30 kpc. This is not too different from, but somewhat smaller
than the measured offsets between fitted positions and literature
X-ray centres or literature positions of cD galaxies as given in
Tab. 2.  Except for the clusters A0514, A3009 and A3128, which are
either double-peaked or irregular, and apart from A3825 all offsets
are smaller than 100 kpc, and the biweight average measured offset
between fitted positions and X-ray or cD positions is 51 kpc.

\begin{table} 
\caption[]{Literature positions for 17 clusters with
either X-ray or cD centre position. Col.(1) cluster name, cols.(2) and
(3) literature centre, col.(4) distance between literature and fitted
centre in kpc, col.(5) type of literature centre (X or cD)}
\begin{flushleft} 
{\scriptsize 
\begin{tabular}{crrrr} 
\hline
\noalign{\smallskip} 
ACO & \multicolumn{1}{c}{$\alpha $} &
      \multicolumn{1}{c}{$\delta $} & dist. & type \\
 &    \multicolumn{2}{c}{2000.0} & (kpc) & \\ 
\noalign{\smallskip}
\hline
\noalign{\smallskip}
0119 & 00:56:17.80 & -01:15:23 &  27 & X  \\ 
0151 & 01:08:50.59 & -15:24:26 &  22 & cD \\ 
0168 & 01:15:08.44 &  00:17:11 &  88 & X  \\
0514 & 04:48:30.35 & -20:32:09 & 413 & X  \\ 
1069 & 10:39:44.29 & -08:41:25 &  58 & X  \\
2426 & 22:14:31.45 & -10:22:27 &  87 & X  \\ 
2734 & 00:11:21.56 & -28:51:15 &  22 & cD \\ 
2911 & 01:26:05.51 & -37:57:54 &  35 & cD \\
2923 & 01:32:21.40 & -31:05:31 &  88 & cD \\ 
3009 & 02:22:06.88 & -48:33:49 & 664 & cD \\ 
3112 & 03:17:56.95 & -44:13:59 &  47 & X  \\
3128 & 03:30:50.85 & -52:30:31 & 485 & X  \\ 
3158 & 03:42:56.64 & -53:38:04 &  63 & X  \\ 
3528 & 12:54:23.54 & -29:01:24 &  34 & X  \\
3806 & 21:46:21.77 & -57:17:11 &  19 & cD \\ 
3825 & 21:58:40.28 & -60:19:57 & 162 & X  \\ 
3827 & 22:01:57.88 & -59:56:18 &  52 & X  \\
\hline	   
\end{tabular}
}
\normalsize
\end{flushleft}
\end{table}

In Tab. 3 and Fig.~3, we show a comparison with the more homogeneous ROSAT 
sample of X-ray centres (preliminary centres kindly provided by H. Bohringer). 
The biweight average offset between fitted positions and X-ray centres is 78 
kpc. If we ignore the clusters with an atypical offset of more than 150 kpc, 
we obtain 69 kpc. The clusters that we considered atypical are A0380 which 
is a clear 
double-peaked cluster, A3809 which is irregular, A2871 which exhibits clearly 
a very atypical difference of 449 kpc, A0524 and A2799. We have checked the 
internal accuracy of the ROSAT centres by using the X-ray map of A0119 (kindly 
provided by D. Neumann). We estimate that this error is about 20'' (11 kpc) 
for A0119. However, we deal we a very regular cluster and the error is probably 
underestimated compared to the other clusters. Another way to estimate the 
global error of the ROSAT centres is to compare with other literature 
positions (X-ray or cD). We find a mean offset of 31 kpc (see Tab. 3), 
greater than the 11 kpc obtained for A0119.

\begin{figure} 
\vbox 
{\psfig{file=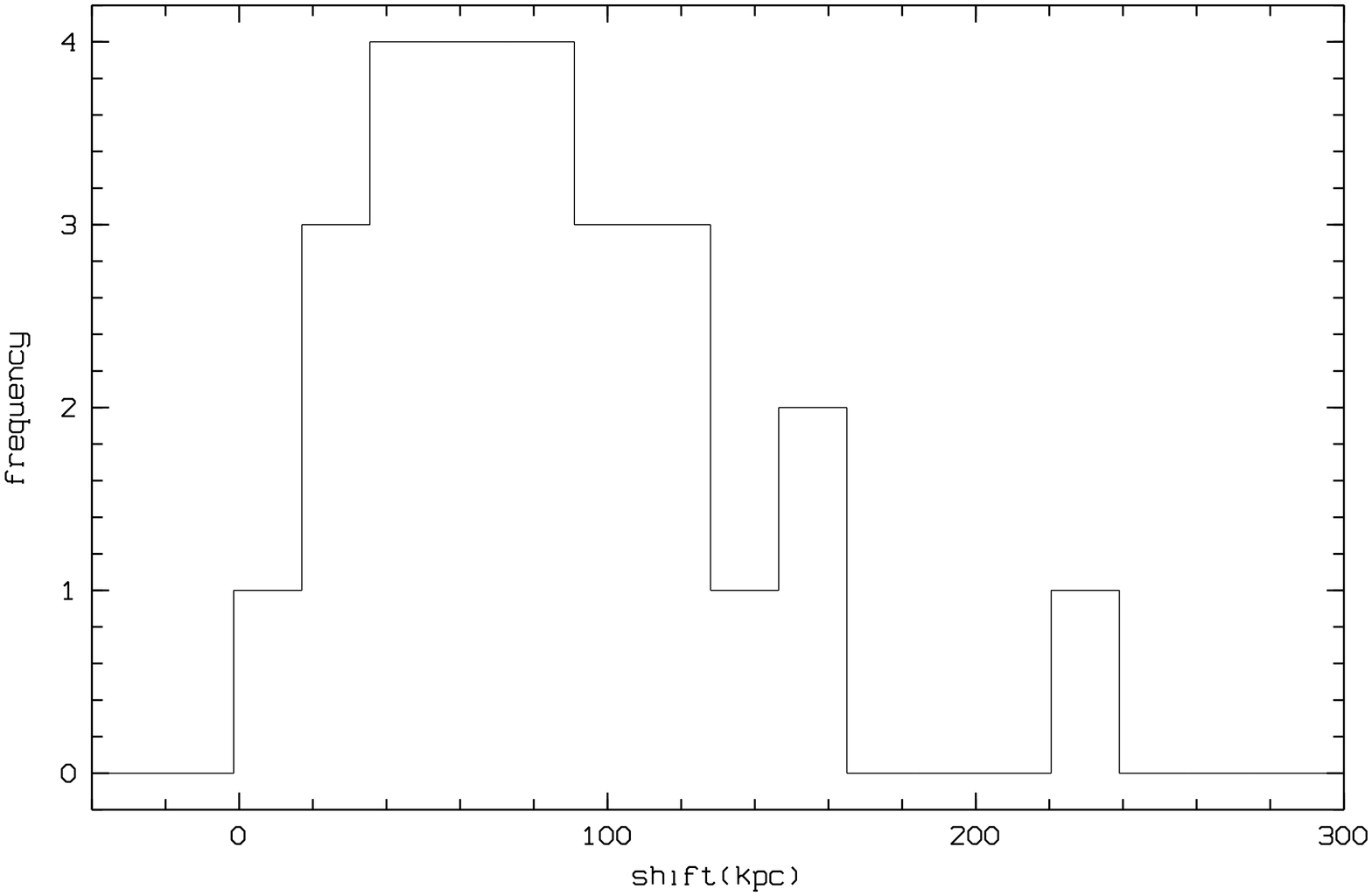,width=9.0cm,angle=270}}
\caption[]{The histogram of the offsets between the X-Ray ROSAT and the fitted 
centres. The two clearly atypic values for A0380 and A2871 are not shown.}  
\end{figure}

\begin{table} 
\caption[]{ROSAT X-ray centre positions for 28 clusters.
Col.(1) cluster name, cols.(2) and (3) ROSAT centre, col.(4) distance
between ROSAT and fitted centre in kpc, col.(5) distance between ROSAT
and previous literature centre in kpc (when available).}
\begin{flushleft} 
{\scriptsize 
\begin{tabular}{crrrc} 
\hline
\noalign{\smallskip} 
  ACO & \multicolumn{1}{c}{$\alpha $} &
        \multicolumn{1}{c}{$\delta $} & dist. & dist. ROSAT/Lit \\
      &   \multicolumn{2}{c}{2000.0} & (kpc) & (kpc) \\ 
\noalign{\smallskip}
\hline
\noalign{\smallskip}
0013 & 00:13:38.23 & -19:30:08 &  75 &    \\ 
0119 & 00:56:14.60 & -01:15:44 &  47 & 40 \\ 
0151 & 01:08:50.47 & -15:24:33 &  18 &  4 \\ 
0367 & 02:36:39.34 & -19:23:09 &  94 &    \\
0380 & 02:44:05.69 & -26:11:17 & 501 &    \\ 
0524 & 04:57:57.19 & -19:43:58 & 150 &    \\ 
0978 & 10:20:28.72 & -06:31:14 &  75 &    \\ 
1069 & 10:39:44.81 & -08:41:01 &  42 & 22 \\
2426 & 22:14:32.38 & -10:22:12 &  61 & 25 \\ 
2717 & 00:03:11.14 & -35:55:21 &  74 &    \\ 
2734 & 00:11:20.28 & -28:51:31 &  44 & 21 \\ 
2755 & 00:17:34.90 & -35:11:09 &  97 &    \\ 
2764 & 00:20:34.10 & -49:13:40 &  72 &    \\ 
2799 & 00:37:29.47 & -39:06:18 & 152 &    \\ 
2871 & 01:07:49.37 & -36:43:44 & 449 &    \\ 
2911 & 01:26:04.56 & -37:58:36 &  71 & 38 \\
3093 & 03:10:55.15 & -47:24:31 &  40 &    \\ 
3112 & 03:17:58.27 & -44:14:16 &   7 & 25 \\
3122 & 03:22:18.19 & -41:21:29 & 136 &    \\ 
3158 & 03:42:54.82 & -53:38:08 & 120 & 22 \\ 
3194 & 03:59:08.16 & -30:11:36 &  25 &    \\ 
3341 & 05:25:32.66 & -31:35:56 &  24 &    \\ 
3744 & 21:07:13.49 & -25:26:54 & 122 &    \\
3806 & 21:46:27.00 & -57:16:47 &  60 & 81 \\ 
3809 & 21:46:57.72 & -43:54:33 & 224 &    \\ 
3822 & 21:54:09.62 & -57:51:15 & 105 &    \\ 
3825 & 21:58:27.26 & -60:23:56 & 113 &    \\ 
3827 & 22:01:55.90 & -59:56:57 &  76 &    \\
\hline	   
\end{tabular}
}
\normalsize
\end{flushleft}
\end{table}

The values of the ellipticities are mostly quite small, and even
indistinghuisable from 0.0 for quite a few of the clusters. This might
seem to be in contradiction to other results on the elongation of
clusters (e.g. Plionis et al. 1991, and de Theije et al. 1995).
However, it must be appreciated that our ellipticities are {\em
apparent} ellipticities, i.e. not corrected for the effect of the
aperture, and that they refer to the central regions only (about 5 King
core radii: i.e. a radius of slightly more than 500 kpc).

\subsection{The values of r$_c$, $\beta$ and $\sigma_b$}

Using the fitted values of x$_0$, y$_0$, e, and $\phi$ we subsequently
made MLM fits for r$_c$, $\beta$ and $\sigma_b$, for the 62 clusters
in Tab. 1, for each of the 4 profiles, for which the solution of
x$_0$, y$_0$, e, and $\phi$ converged.

In Tabs. 4 to 7, we give the values of r$_c$, $\beta$ and $\sigma_b$
for all clusters in the sample of 62, i.e. for all clusters with a
reliable centre position, for each of the model profiles.  For some of
the clusters, one or more of the MLM fits did not converge for one or
more of the parameters. In those cases we give the SIMPLEX values
without error estimates. The error estimates obtained for each of the
3 parameters, as given by MIGRAD are also given in Tabs.  4 to 7, in
which we also give the values of --2ln(L). As is well known,
the magnitude of this parameter (the maximum likelihood for the
particular model profile with the best-fit parameters) cannot be used
to make a statement about the absolute probability that a given
observation is indeed induced by the underlying continuous probability
function described by the particular profile.

As a check on the `meaning' of the error estimates from MIGRAD, we
have generated about 150000 artificial clusters, for which the total
number of galaxies, the ratio between the number of galaxies in the
cluster and in the background, and r$_c$- and $\beta$-values for the 4
model profiles were chosen in the ranges spanned by the
observations. To each of the artificial clusters a MLM fit was made in
the same manner as for the observed clusters. The errors in the
parameters of the fits to the artifical clusters depend somewhat on
the type of profile and on the assumed parameters, but globally, the
results can be summarized as follows.

The background density is recovered with an average accuracy of about
25\% (with actual errors between about 10 and 40\%), with a slightly
better result for the King and Hubble profiles than for the NFW and de
Vaucouleurs profiles (which is not surprising). Almost identical
percentage errors are found for r$_c$. In general, $\beta$ is slightly
better known, with an average error of about 15\% (with actual errors
between about 5 and 25\%). As can be seen from Tabs. 4 to 7, the
errors given by MIGRAD are essentially always in the ranges found for
the artificial clusters.

\begin{table} 
\caption[]{Fitted parameters for the King models and contrast (last
column).}  
\begin{flushleft} 
\small 
{\scriptsize 
\begin{tabular}{crrrrr} \hline
\noalign{\smallskip} ACO & $r_c$ (kpc) & $\beta$ & $\sigma _b$ &
--2ln(L) & C \\ 
\noalign{\smallskip} \hline \noalign{\smallskip} 
0013 & 68$\pm$18 & 0.97$\pm$0.11 & 8.7$\pm$1.0 & 2397.90 & 0.504 \\ 
0087 &118$\pm$12 & 1.06$\pm$0.15 & 3.0$\pm$0.6 & 1624.15 & 0.346 \\ 
0118 & 118$\pm$44 & 1.04$\pm$0.07 & 9.1$\pm$1.3 & 1622.29 & 0.262 \\ 
0119 & 55$\pm$14 & 1.07$\pm$0.10 & 4.5$\pm$0.2 & 6963.61 & 0.510 \\ 
0151 & 56$\pm$ 9 & 1.18$\pm$0.15 & 4.6$\pm$0.4 & 3999.36 & 0.646 \\ 
0168 & 161$\pm$42 & 1.01$\pm$0.28 & 4.2$\pm$0.9 & 2179.73 & 0.454 \\ 
0229 & 43 & 0.97 & 9.0 & 839.70 & 0.784 \\ 
0367 & 128$\pm$21 & 1.23$\pm$0.13 & 3.4$\pm$0.7 & 2043.15 & 0.537 \\ 
0380 & 496$\pm$86 & 1.02$\pm$0.22 & 2.0$\pm$0.2 & 1753.21 & 0.976 \\ 
0514 & 90$\pm$25 & 1.01$\pm$0.14 & 5.4$\pm$1.1 & 2358.88 & 0.547 \\ 
0524 & 95$\pm$23 & 1.04$\pm$0.09 & 1.1$\pm$0.5 & 1495.57 & 0.823 \\ 
0978 & 44$\pm$12 & 0.94$\pm$0.15 & 3.5$\pm$0.3 & 3564.50 & 0.648 \\ 
1069 & 219$\pm$65 & 1.03$\pm$0.29 & 3.0$\pm$1.0 & 2704.82 & 0.465 \\ 
2353 & 163$\pm$40 & 0.96$\pm$0.12 & 2.9$\pm$1.6 & 1839.96 & 0.519 \\ 
2362 & 110$\pm$32 & 1.01$\pm$0.28 & 4.6$\pm$1.2 & 1636.26 & 0.256 \\ 
2383 & 86$\pm$19 & 0.95$\pm$0.12 & 3.1$\pm$0.6 & 2304.33 & 0.517 \\ 
2426 & 135$\pm$26 & 0.84$\pm$0.08 & 5.7$\pm$0.9 & 5044.65 & 0.579 \\ 
2436 & 79$\pm$37 & 1.00$\pm$0.19 & 7.0$\pm$0.9 & 1195.09 & 0.318 \\ 
2480 & 101$\pm$28 & 1.09$\pm$0.33 & 4.8$\pm$0.9 & 1434.17 & 0.309 \\ 
2644 & 76$\pm$22 & 1.00$\pm$0.14 & 3.3$\pm$0.8 & 1249.19 & 0.846 \\ 
2715 & 236 & 1.00 & 3.0 & 2639.96 & 0.594 \\ 
2717 & 140$\pm$40 & 0.99$\pm$0.11 & 3.2$\pm$0.4 & 4859.57 & 0.470 \\ 
2721 & 286$\pm$34 & 0.96$\pm$0.08 & 3.5$\pm$1.0 & 6146.76 & 0.696 \\ 
2734 & 105$\pm$22 & 1.00$\pm$0.11 & 3.6$\pm$0.4 & 5277.62 & 0.567 \\ 
2755 & 49$\pm$10 & 1.01$\pm$0.07 & 9.5$\pm$0.9 & 3842.01 & 0.754 \\ 
2764 & 101$\pm$22 & 1.00$\pm$0.13 & 5.6$\pm$0.9 & 3514.23 & 0.511 \\ 
2765 & 66$\pm$22 & 0.98$\pm$0.19 & 2.9$\pm$0.4 & 1757.98 & 0.661 \\ 
2778 & 98$\pm$26 & 1.01$\pm$0.13 & 4.1$\pm$0.7 & 1746.72 & 0.577 \\ 
2799 & 46$\pm$11 & 0.94$\pm$0.33 & 4.5$\pm$0.5 & 2022.80 & 0.514 \\ 
2800 & 99$\pm$25 & 1.01$\pm$0.14 & 4.0$\pm$0.6 & 2050.76 & 0.282 \\ 
2854 & 67$\pm$17 & 1.01$\pm$0.10 & 2.6$\pm$0.7 & 1517.14 & 0.643 \\ 
2871 & 139$\pm$30 & 1.06$\pm$0.22 & 2.1$\pm$0.8 & 2306.66 & 0.796 \\ 
2911 & 109$\pm$28 & 1.13$\pm$0.30 & 8.8$\pm$0.9 & 2842.87 & 0.268 \\ 
2923 & 142$\pm$19 & 1.33$\pm$0.18 & 2.1$\pm$0.6 & 1110.31 & 0.914 \\
3009 & 45$\pm$33 & 0.99$\pm$0.20 & 7.4$\pm$0.5 & 1623.65 & 0.230 \\ 
3093 & 69$\pm$21 & 1.00$\pm$0.12 & 5.7$\pm$1.6 & 1544.33 & 0.457 \\ 
3094 & 127$\pm$30 & 1.01$\pm$0.16 & 5.7$\pm$0.8 & 2907.21 & 0.458 \\ 
3111 & 99$\pm$26 & 1.00$\pm$0.03 & 4.2$\pm$0.4 & 5842.78 & 0.591 \\ 
3112 & 229$\pm$29 & 1.02$\pm$0.28 & 3.7$\pm$1.8 & 2277.64 & 0.502 \\ 
3122 & 150$\pm$20 & 0.97$\pm$0.07 & 3.2$\pm$0.5 & 6535.82 & 0.564 \\ 
3128 & 362$\pm$29 & 1.08$\pm$0.08 & 3.1$\pm$0.5 & 12479.20 & 0.500 \\ 
3141 & 176$\pm$32 & 0.97$\pm$0.10 & 1.3$\pm$0.7 & 1646.93 & 0.870 \\ 
3158 & 102$\pm$15 & 0.95$\pm$0.07 & 4.5$\pm$0.7 & 7446.85 & 0.639 \\ 
3194 & 54$\pm$ 7 & 1.02$\pm$0.23 & 7.0$\pm$1.3 & 1306.30 & 0.601 \\ 
3202 & 64$\pm$21 & 0.99$\pm$0.14 & 3.6$\pm$1.0 & 1652.64 & 0.657 \\ 
3223 & 154$\pm$39 & 1.01$\pm$0.12 & 3.0$\pm$1.6 & 2632.48 & 0.674 \\ 
3341 & 51$\pm$12 & 0.95$\pm$0.13 & 4.0$\pm$0.8 & 2239.58 & 0.466 \\ 
3528 & 135$\pm$35 & 1.06$\pm$0.26 & 9.0$\pm$3.0 & 3130.11 & 0.311 \\ 
3705 & 99$\pm$28 & 0.98$\pm$0.20 & 6.7$\pm$2.3 & 999.64 & 0.639 \\ 
3733 & 37$\pm$11 & 1.07$\pm$0.15 & 3.1$\pm$0.9 & 1183.78 & 0.659 \\ 
3744 & 79$\pm$19 & 1.10$\pm$0.23 & 4.2$\pm$0.6 & 2339.98 & 0.188 \\ 
3764 & 59$\pm$16 & 1.31$\pm$0.18 & 5.5$\pm$0.7 & 1754.89 & 0.556 \\ 
3781 & 232$\pm$71 & 1.01$\pm$0.21 & 2.6$\pm$1.5 & 1580.22 & 0.520 \\ 
3806 & 224$\pm$30 & 1.01$\pm$0.12 & 1.1$\pm$1.5 & 2953.74 & 0.688 \\ 
3806 & 106$\pm$33 & 1.00$\pm$0.23 & 14.$\pm$13. & 3533.92 & 0.669 \\ 
3809 & 365$\pm$58 & 1.00$\pm$0.23 & 2.5$\pm$1.2 & 5251.67 & 0.708 \\ 
3822 & 242$\pm$32 & 1.04$\pm$0.11 & 8.4$\pm$1.1 & 7313.07 & 0.282 \\ 
3825 & 108$\pm$26 & 0.97$\pm$0.12 & 8.8$\pm$1.0 & 3802.11 & 0.245 \\ 
3827 & 102$\pm$17 & 1.00$\pm$0.09 & 5.0$\pm$1.6 & 2646.64 & 0.707 \\ 
3897 & 76 & 1.01 & 4.0 & 1028.79 & 0.498 \\ 
\hline 
\end{tabular} } 
\normalsize
\end{flushleft} 
\end{table}

\begin{table} 
\caption[]{Fitted parameters for the Hubble models.}
\begin{flushleft} 
\small {\scriptsize 
\begin{tabular}{crrrr} \hline
\noalign{\smallskip} ACO & $r_c$ (kpc) & $\beta$ & $\sigma _b$ &
 --2ln(L) \\ \noalign{\smallskip} \hline
\noalign{\smallskip}
0013  & 122$\pm$14 & 0.97$\pm$0.11 &  5.3$\pm$1.3 & 2399.12 \\ 
0087  & 183$\pm$51 & 1.02$\pm$0.25 &  1.5$\pm$0.6 & 1624.63 \\ 
0118  & 135$\pm$33 & 1.01$\pm$0.25 & 10.0$\pm$1.4 & 1623.69 \\ 
0119  &  73$\pm$10 & 1.14$\pm$0.10 &  4.0$\pm$0.9 & 6965.28 \\ 
0151  &  94$\pm$16 & 1.19$\pm$0.17 &  7.5$\pm$1.0 & 4001.52 \\ 
0168  & 178$\pm$28 & 0.98$\pm$0.14 &  1.7$\pm$1.9 & 2182.01 \\ 
0367  & 162$\pm$26 & 1.23$\pm$0.14 &  2.6$\pm$1.2 & 2043.29 \\ 
0380  & 556$\pm$95 & 0.98$\pm$0.24 &  0.6$\pm$2.0 & 1754.17 \\ 
0514  & 117$\pm$33 & 1.06$\pm$0.15 &  6.3$\pm$2.1 & 2359.08 \\ 
1069  & 411$\pm$65 & 1.03$\pm$0.19 &  1.1$\pm$1.3 & 2705.29 \\ 
2353  & 198$\pm$32 & 0.93$\pm$0.04 &  0.8$\pm$1.2 & 1838.87 \\ 
2362  & 257$\pm$28 & 1.05$\pm$0.13 &  1.1$\pm$0.7 & 1636.31 \\ 
2383  &  88$\pm$33 & 0.99$\pm$0.21 &  2.1$\pm$1.6 & 2303.26 \\ 
2426  & 253$\pm$22 & 1.01$\pm$0.36 &  7.0$\pm$0.9 & 5048.38 \\ 
2480  & 272        & 1.01          &  3.5         & 1435.84 \\ 
2644  & 118$\pm$33 & 0.97$\pm$0.35 &  1.3$\pm$0.5 & 1251.32 \\ 
2715  & 100$\pm$25 & 0.99$\pm$0.25 &  5.8$\pm$1.3 & 2642.28 \\ 
2717  & 156$\pm$23 & 1.00$\pm$0.29 &  2.7$\pm$0.9 & 4859.71 \\ 
2721  & 264$\pm$42 & 0.97$\pm$0.46 &  5.0$\pm$1.5 & 6148.75 \\ 
2734  & 137        & 0.99$\pm$0.14 &  3.7         & 5277.09 \\ 
2755  &  54$\pm$ 7 & 0.99$\pm$0.20 &  7.1$\pm$1.5 & 3842.10 \\ 
2764  & 126$\pm$25 & 1.01$\pm$0.12 &  6.6$\pm$1.9 & 3514.66 \\ 
2799  &  48$\pm$15 & 1.02$\pm$0.10 &  5.4$\pm$0.9 & 2024.67 \\ 
2800  &  87$\pm$26 & 1.01$\pm$0.25 &  4.6         & 2052.23 \\ 
2854  &  70$\pm$11 & 1.13$\pm$0.16 &  2.0$\pm$0.9 & 1517.04 \\ 
2871  & 142$\pm$17 & 0.99$\pm$0.14 &  0.9$\pm$0.5 & 2308.41 \\ 
2923  &  96$\pm$19 & 1.01$\pm$0.11 &  0.2$\pm$1.8 & 1114.55 \\ 
3094  & 300        & 1.07          &  3.4         & 2908.45 \\ 
3111  & 180$\pm$23 & 1.01$\pm$0.15 &  0.1$\pm$0.7 & 5846.81 \\ 
3112  & 248$\pm$65 & 1.01$\pm$0.18 &  5.0$\pm$0.9 & 2277.91 \\ 
3122  & 319         & 1.19           &  1.6       & 6534.76 \\ 
3128  & 520$\pm$39 & 1.01$\pm$0.25 &  1.1$\pm$1.2 & 12480.05 \\ 
3158  & 104$\pm$12 & 0.87$\pm$0.18 &  4.2$\pm$1.3 & 7445.38 \\ 
3202  & 132$\pm$39 & 1.18$\pm$0.18 &  3.7$\pm$1.5 & 1653.53 \\ 
3223  & 211$\pm$24 & 1.01$\pm$0.20 &  2.0$\pm$1.0 & 2632.55 \\ 
3341  & 133$\pm$19 & 1.11$\pm$0.19 &  1.6$\pm$1.1 & 2240.54 \\ 
3733  &  84$\pm$15 & 1.01$\pm$0.11 &  3.1$\pm$1.5 & 1185.51 \\ 
3744  & 155$\pm$22 & 0.96$\pm$0.20 &  1.1$\pm$0.5 & 2340.79 \\ 
3764  &  97$\pm$12 & 1.13$\pm$0.29 &  6.7$\pm$1.0 & 1760.46 \\ 
3781  & 297$\pm$75 & 1.02$\pm$0.30 &  1.1$\pm$0.8 & 1579.93 \\ 
3806  & 200$\pm$38 & 1.01$\pm$0.19 & 10.5$\pm$2.2 & 2959.95 \\ 
3806  & 284$\pm$31 & 1.04$\pm$0.13 &  3.0$\pm$6.9 & 3535.88 \\ 
3822  & 378$\pm$61 & 1.06$\pm$0.23 &  6.6$\pm$0.9 & 7313.49 \\ 
3825  & 244$\pm$50 & 0.97$\pm$0.10 &  7.0$\pm$2.0 & 3803.01 \\ 
3827  & 119$\pm$13 & 1.01$\pm$0.08 &  4.3$\pm$0.7 & 2646.25 \\ 
\hline	   
\end{tabular}
}
\normalsize
\end{flushleft}
\end{table}

\begin{table}
\caption[]{Fitted parameters for the NFW models.}
\begin{flushleft}
\small
{\scriptsize
\begin{tabular}{crrrr}
\hline
\noalign{\smallskip}
ACO & $r_c$ (kpc) & $\beta$ & $\sigma_b$ & --2ln(L) \\ 
\noalign{\smallskip}
\hline
\noalign{\smallskip}
0013 & 128$\pm$63 & 0.64$\pm$0.07 & 8.4$\pm$1.1 & 2399.69 \\ 
0087 & 324$\pm$35 & 0.68$\pm$0.13 & 2.8$\pm$0.7 & 1625.99 \\ 
0118 & 169$\pm$134 & 0.58$\pm$0.13 & 10.0$\pm$1.8 & 1624.48 \\ 
0119 & 146$\pm$25 & 0.70$\pm$0.10 & 4.0$\pm$0.6 & 6965.88 \\ 
0151 & 112$\pm$28 & 0.68$\pm$0.14 & 4.5$\pm$1.1 & 4000.93 \\ 
0168 & 210$\pm$39 & 0.56$\pm$0.10 & 1.0$\pm$0.4 & 2179.72 \\ 
0367 & 286$\pm$44 & 0.62$\pm$0.11 & 1.4$\pm$0.9 & 2043.92 \\ 
0524 & 141$\pm$55 & 0.62$\pm$0.05 & 0.2$\pm$0.6 & 1497.56 \\ 
0978 &  94$\pm$50 & 0.64$\pm$0.07 & 3.3$\pm$0.3 & 3564.57 \\ 
1069 & 281$\pm$34 & 0.61$\pm$0.09 & 4.9$\pm$1.1 & 2707.30 \\ 
2353 & 327$\pm$175 & 0.59$\pm$0.07 & 2.6$\pm$1.0 & 1838.57 \\ 
2362 & 248$\pm$49 & 0.55$\pm$0.10 & 3.0$\pm$1.0 & 1636.39 \\ 
2383 & 143$\pm$86 & 0.66$\pm$0.08 & 3.3$\pm$0.6 & 2301.49 \\ 
2426 & 458$\pm$185 & 0.63$\pm$0.07 & 5.0$\pm$1.1 & 5044.38 \\ 
2436 & 193$\pm$192 & 0.66$\pm$0.17 & 7.0$\pm$1.2 & 1195.58 \\ 
2480 & 430$\pm$50 & 0.60$\pm$0.09 & 4.0$\pm$0.8 & 1436.00 \\ 
2644 & 106$\pm$37 & 0.55$\pm$0.06 & 2.5$\pm$0.4 & 1251.34 \\ 
2717 & 458$\pm$238 & 0.58$\pm$0.06 & 1.8$\pm$0.6 & 4859.71 \\ 
2721 & 649         & 0.55          & 2.0         & 6150.09 \\ 
2734 & 253$\pm$35 & 0.53$\pm$0.11 & 3.0$\pm$1.2 & 5276.62 \\ 
2755 & 173$\pm$63 & 0.67$\pm$0.07 & 5.9$\pm$1.1 & 3841.34 \\ 
2764 & 356$\pm$46 & 0.61$\pm$0.09 & 5.5$\pm$0.6 & 3515.77 \\ 
2765 & 147$\pm$96 & 0.55$\pm$0.08 & 2.0$\pm$0.6 & 1757.89 \\ 
2778 & 390$\pm$239 & 0.70$\pm$0.09 & 2.8$\pm$1.0 & 1746.35 \\ 
2799 & 296$\pm$52 & 0.67$\pm$0.12 & 2.7$\pm$0.6 & 2022.63 \\ 
2800 & 266$\pm$43 & 0.59$\pm$0.09 & 2.7$\pm$1.0 & 2052.00 \\ 
2854 & 150$\pm$35 & 0.69$\pm$0.07 & 2.1$\pm$0.7 & 1516.08 \\ 
2871 & 226$\pm$75 & 0.57$\pm$0.05 & 0.5$\pm$1.0 & 2310.69 \\ 
2911 & 392$\pm$48 & 0.59$\pm$0.05 & 6.3$\pm$0.3 & 2844.33 \\ 
2923 & 123$\pm$43 & 0.63$\pm$0.06 & 0.2$\pm$0.6 & 1115.98 \\ 
3009 & 95         & 0.61          & 7.4         & 1623.92 \\ 
3093 & 362        & 0.62          & 3.0         & 1543.66 \\ 
3094 & 247$\pm$121 & 0.57$\pm$0.09 & 6.0$\pm$1.0 & 2909.27 \\ 
3111 & 144$\pm$66 & 0.51$\pm$0.04 & 4.1$\pm$0.5 & 5848.95 \\ 
3112 & 455$\pm$57 & 0.57$\pm$0.10 & 6.6$\pm$1.7 & 2279.49 \\ 
3122 & 339        & 0.53          & 7.1         & 6535.54 \\ 
3128 & 966        & 0.59          & 2.7         & 12484.70 \\ 
3158 & 430$\pm$151 & 0.63$\pm$0.07 & 3.3$\pm$1.1 & 7442.27 \\ 
3194 & 100$\pm$54 & 0.65$\pm$0.09 & 6.5$\pm$1.6 & 1308.37 \\ 
3202 & 348$\pm$41 & 0.69$\pm$0.06 & 4.8$\pm$0.8 & 1652.14 \\ 
3341 & 275$\pm$190 & 0.61$\pm$0.04 & 3.0$\pm$1.0 & 2240.44 \\ 
3528 & 168         & 0.57          & 2.3        & 3143.42 \\
3705 & 340$\pm$205 & 0.66$\pm$0.08 & 3.8$\pm$3.3 & 995.70 \\ 
3733 & 99$\pm$29 & 0.64$\pm$0.09 & 3.8$\pm$1.0 & 1185.51 \\ 
3744 & 202$\pm$133 & 0.67$\pm$0.12 & 3.9$\pm$0.7 & 2342.55 \\ 
3764 & 100$\pm$39 & 0.68$\pm$0.07 & 5.4$\pm$1.0 & 1757.60 \\ 
3781 & 849$\pm$598 & 0.69$\pm$0.15 & 3.5$\pm$1.1 & 1578.26 \\ 
3806 & 671$\pm$492 & 0.55$\pm$0.11 & 10.0$\pm$2.6 & 3534.19 \\ 
3809 & 541$\pm$334 & 0.54$\pm$0.09 & 5.2$\pm$0.9 & 5254.24 \\
3827 & 181$\pm$81 & 0.61$\pm$0.05 & 4.4$\pm$2.4 & 2647.14 \\ 
\hline	   
\end{tabular}
}
\normalsize
\end{flushleft}
\end{table}

\begin{table}
\caption[]{Fitted parameters for the de Vaucouleurs 
models.}
\begin{flushleft}
\small
{\scriptsize
\begin{tabular}{crrrr}
\hline
\noalign{\smallskip}
ACO & $r_c$ (kpc) & $\beta$ & $\sigma_b$ & --2ln(L) \\ 
\noalign{\smallskip}
\hline
\noalign{\smallskip}
0013  & 1441$\pm$379 & 7.46$\pm$0.87 & 5.4$\pm$1.4 & 2399.25 \\ 
0087  & 691           & 7.32           & 3.3          & 1626.20 \\ 
0118  & 2961$\pm$1778 & 7.42$\pm$1.59 & 6.5$\pm$2.5 & 1623.77 \\ 
0119  &  861$\pm$210 & 7.60$\pm$0.76 & 3.8$\pm$0.3 & 6969.06 \\ 
0151  &  573$\pm$260 & 7.78$\pm$0.94 & 3.8$\pm$ 0.5 & 4001.32 \\ 
0168  & 1031$\pm$104 & 7.45$\pm$0.60 & 5.0$\pm$0.7 & 2179.82 \\ 
0367  & 1535$\pm$580 & 7.70$\pm$0.83 & 1.4$\pm$0.8 & 2043.87 \\ 
0380  & 4762          & 7.03           & 2.6          & 1755.37 \\ 
0514  & 1752$\pm$304 & 7.60$\pm$0.78 & 6.4$\pm$1.5 & 2359.24 \\ 
0524  & 1146$\pm$403 & 8.92$\pm$0.74 & 0.9$\pm$0.5  & 1497.69 \\ 
0978  &  458$\pm$200 & 7.82$\pm$0.86 &  3.4$\pm$0.3 & 3564.55 \\ 
1069  & 1365$\pm$411 & 7.51$\pm$0.89 & 4.5$\pm$0.9 & 2708.09 \\ 
2353  & 1197$\pm$582 & 7.06$\pm$0.90 & 3.0$\pm$0.9 & 1837.85 \\ 
2362  & 1721$\pm$304 & 8.05$\pm$0.92 & 4.1$\pm$0.3 & 1636.56 \\ 
2383  & 1184$\pm$576 & 7.95$\pm$0.96 & 2.7$\pm$0.7 & 2301.50 \\ 
2426  & 1272          & 7.50           & 6.4           & 5045.18 \\ 
2436  & 1438$\pm$1060 & 8.72$\pm$2.17 &  7.0$\pm$1.3 & 1195.54 \\ 
2480  & 1161$\pm$286 & 7.52$\pm$0.58 &  4.6$\pm$1.1 & 1436.05 \\ 
2644  & 1074$\pm$311 & 7.78$\pm$0.79 & 2.5$\pm$0.8 &  1251.41 \\ 
2717  & 1971$\pm$841 & 6.59$\pm$0.72 & 2.0$\pm$0.7 & 4859.83 \\ 
2721  & 2879$\pm$1043 & 6.42$\pm$0.51 & 2.3$\pm$1.3 & 6149.83 \\ 
2734  & 1932$\pm$298  & 7.89$\pm$0.70 &  3.6$\pm$0.5 & 5277.25 \\ 
2755  & 2077$\pm$655 & 9.24$\pm$0.69 & 4.0$\pm$1.3 & 3842.03 \\ 
2764  & 3165$\pm$379 & 7.87$\pm$0.55 & 5.0$\pm$1.3 & 3515.44 \\ 
2765  & 1460$\pm$840 & 7.62$\pm$1.11 &  2.0$\pm$0.6 & 1757.91 \\ 
2778  & 1263          & 7.51           & 2.8            & 1746.40 \\ 
2799  &  787$\pm$255 & 7.56$\pm$0.33 & 3.3$\pm$0.8 & 2022.67 \\ 
2800  & 1712$\pm$213 & 7.80$\pm$0.83 &  3.1$\pm$0.8 & 2051.98 \\ 
2854  &  802$\pm$266 & 7.82$\pm$0.72 & 1.7$\pm$0.7 & 1516.30 \\ 
2871  & 1608         & 7.84           & 0.4           & 2310.71 \\ 
2911  & 2213$\pm$266 & 7.18$\pm$0.22 &  6.4$\pm$1.4 & 2844.38 \\ 
2923  &  735$\pm$207 & 8.12$\pm$0.57 & 3.0$\pm$0.3 & 1116.00 \\ 
3009  & 2302         & 7.61           & 6.9          & 1623.90 \\ 
3093  & 1028         & 6.99           & 3.6          & 1543.57 \\ 
3094  & 1406         & 7.52           & 5.4          & 2909.06 \\ 
3111  & 1620$\pm$352 & 7.09$\pm$0.43 & 4.0$\pm$0.5 & 5848.64 \\ 
3112  & 2914$\pm$309 & 7.69$\pm$0.78 & 6.2$\pm$1.7 & 2279.38 \\ 
3122  & 1655$\pm$338 & 7.68$\pm$0.50 &  3.1$\pm$0.6 & 6537.58 \\ 
3128  & 1055$\pm$202 & 7.51$\pm$0.57 & 5.8$\pm$0.4 & 12494.90 \\ 
3141  & 1971$\pm$640 & 7.52$\pm$0.62 &  1.9$\pm$3.1 & 1672.27 \\ 
3158  & 1721$\pm$255 & 7.51$\pm$0.58 & 4.1$\pm$0.3 & 7443.36 \\ 
3194  & 1278         & 7.71           & 4.3          & 1308.64 \\ 
3202  & 1454$\pm$289 & 7.66$\pm$0.54 & 4.0$\pm$1.4 & 1652.45 \\ 
3341  & 1290$\pm$768 & 7.69$\pm$1.17 & 2.8$\pm$0.9 & 2240.12 \\
3705  & 1311$\pm$710 & 7.68$\pm$1.04 &  3.9$\pm$3.8  & 996.19 \\ 
3733  & 1454$\pm$158 & 7.83$\pm$0.82 & 2.1$\pm$1.3 & 1185.71 \\ 
3764  &  491$\pm$235 & 7.10$\pm$0.59 & 4.0$\pm$0.3 & 1757.42 \\ 
3781  & 1600$\pm$1226 & 8.01$\pm$1.82 & 4.3$\pm$0.9 & 1578.42 \\ 
3806  & 2162$\pm$956 & 7.08$\pm$1.48 & 1.3$\pm$1.9 & 3534.43 \\ 
3825  & 0.97$\pm$0.10 & 7.51$\pm$0.59 & 7.9$\pm$1.2 & 3804.20 \\ 
3827  & 2292$\pm$327 & 8.34$\pm$0.25 & 3.5$\pm$2.9 & 2646.07 \\ 
\hline	   
\end{tabular}
}
\normalsize
\end{flushleft}
\end{table}

The average values of r$_c$ and the dispersion around the mean are 128
$\pm$ 88, 189 $\pm$ 116, 292 $\pm$ 191 and 1582 $\pm$ 771 kpc, for the
King, Hubble, NFW and de Vaucouleurs profiles respectively. There are
no general relations for the four values of r$_c$ that one obtains by
fitting an arbitrary galaxy distribution with the four model profiles.
However, from linear regressions between individual values we find, in
our dataset, that:

\noindent ${r_c}_K = (0.62\pm 0.07) {r_c}_H + (18\pm 53)$

\noindent ${r_c}_K = (0.28\pm 0.04) {r_c}_{NFW} + (36\pm 51)$

\noindent ${r_c}_K = (0.06\pm 0.01) {r_c}_{deV} + (27\pm 70)$
       
For $\beta$, the average values and dispersions around the mean are
1.02 $\pm$ 0.08, 1.03 $\pm$ 0.07, 0.61 $\pm$ 0.05 and 7.6 $\pm$ 0.5,
for the King, Hubble, NFW and de Vaucouleurs profiles respectively.
Since $\beta$ is closely related to the asymptotic logarithmic slope
of the profile, there are predictions for the relations between the
values of $\beta$ for the four profiles. As can be easily deduced from
the four expressions in $\S$ 3.1, one would expect that $\beta_H$ =
$\beta_K$, $\beta_{NFW}$ = 0.67 $\beta_K$ and $\beta_{deV}$ = 8
$\beta_K$. It is clear that these relations are, to within the errors,
obeyed by the observed average values for the four profiles. From
inspection of Tabs. 4 to 7, it is clear that it is not very meaningful to
make linear regressions between individual $\beta$-values because the
distributions of $\beta$ for each profile type are quite narrow
compared to the estimated errors in the individual $\beta$-values.

The values of the average backgrounds are (4.7$\pm$2.6),
(3.7$\pm$2.6), (4.0$\pm$2.3) and (3.8$\pm$1.7) times 10$^{-5}$
galaxies per square arcsec for the King, Hubble, NFW and de
Vaucouleurs profiles respectively. For this parameter, the prediction
is clearly that it should be identical for the four fits, as is indeed
the case to within the errors. Linear regressions between the
individual values of the background density give:

\noindent ${\sigma_b}_K$ = (0.89$\pm$ 0.12)${\sigma_b}_H$ + (-1.6$\pm$
1.6)$10^{-5}$.

\noindent ${\sigma_b}_K$ = (0.80$\pm$ 0.11)${\sigma_b}_{NFW}$ +
(1.5$\pm$ 1.7)$10^{-5}$

\noindent ${\sigma_b}_K$ = (0.70$\pm$ 0.19)${\sigma_b}_{deV}$ +
(1.9$\pm$ 2.3)$10^{-5}$.

We give also in Tab.~4 the value of the contrast C for each
cluster. This parameter is defined as the fraction of the observed
galaxies in a pencil beam that is really in the cluster.  If $N_{tot}$
is the number of objects of the line of sight and if $N_{bck}$ is the
estimated number of background galaxies, we have

$C=(N_{tot}-N_{bck})/N_{tot}$  

This parameter has been used in paper IV to select the more contrasted
clusters, in order to see a narrower Fundamental Plane. 

\subsection{The distribution of backgrounds on the sky}

The average backgrounds $\sigma _b$ as given in $\S$ 3.2 are in very
good agreement for the four different profiles used, as was to be
expected. The mean of the four values is about 4 10$^{-5}$ gals
arcsec$^{-2}$, and this mean refers to a magnitude limit m$_{b_j}$ of
19.5. For the same limit, Colless (1989) predicted a background of
3 10$^{-5}$ gals arcsec$^{-2}$. Lilly et al. (1995) and Crampton et
al. (1995) found, from CFRS data, 5.6 10$^{-5}$ gals arcsec$^{-2}$ and
Arnouts et al. (1996) and Bellanger et al. (1995), using Sculptor
data, found a background density of 2 10$^{-5}$ gals arcsec$^{-2}$ for
nearly the same limiting magnitude.

These values from the literature vary between about 2 and 6 10$^{-5}$
as do the values of the backgrounds found here. Given the large
variation in our data for what is supposed to be a uniform magnitude
limit, one wonders what is the cause of this variation, and whether
there are correlation between backgrounds found in neighbouring
directions.

In Fig. 4, we show the values of $\sigma _b$ in galactic coordinates
for the clusters with b$\leq$-30$\degr$. In the cone thus defined, we
have backgrounds for two-thirds of the ENACS clusters with z $<$
0.1. We have not included the clusters with b$>$-30$\degr$, as for
those the availability of COSMOS data for the ENACS clusters is much
less complete. Each ENACS pencil beam is indicated by a circle, and
the size of the circle correlates with the value of $\sigma _b$ as
found in the fit of a King profile. The clusters with b$\leq$-30$\degr$
for which no COSMOS data were available, or for which no reliable
centre positions could be determined are indicated by crosses.

\begin{figure} 
\vbox 
{\psfig{file=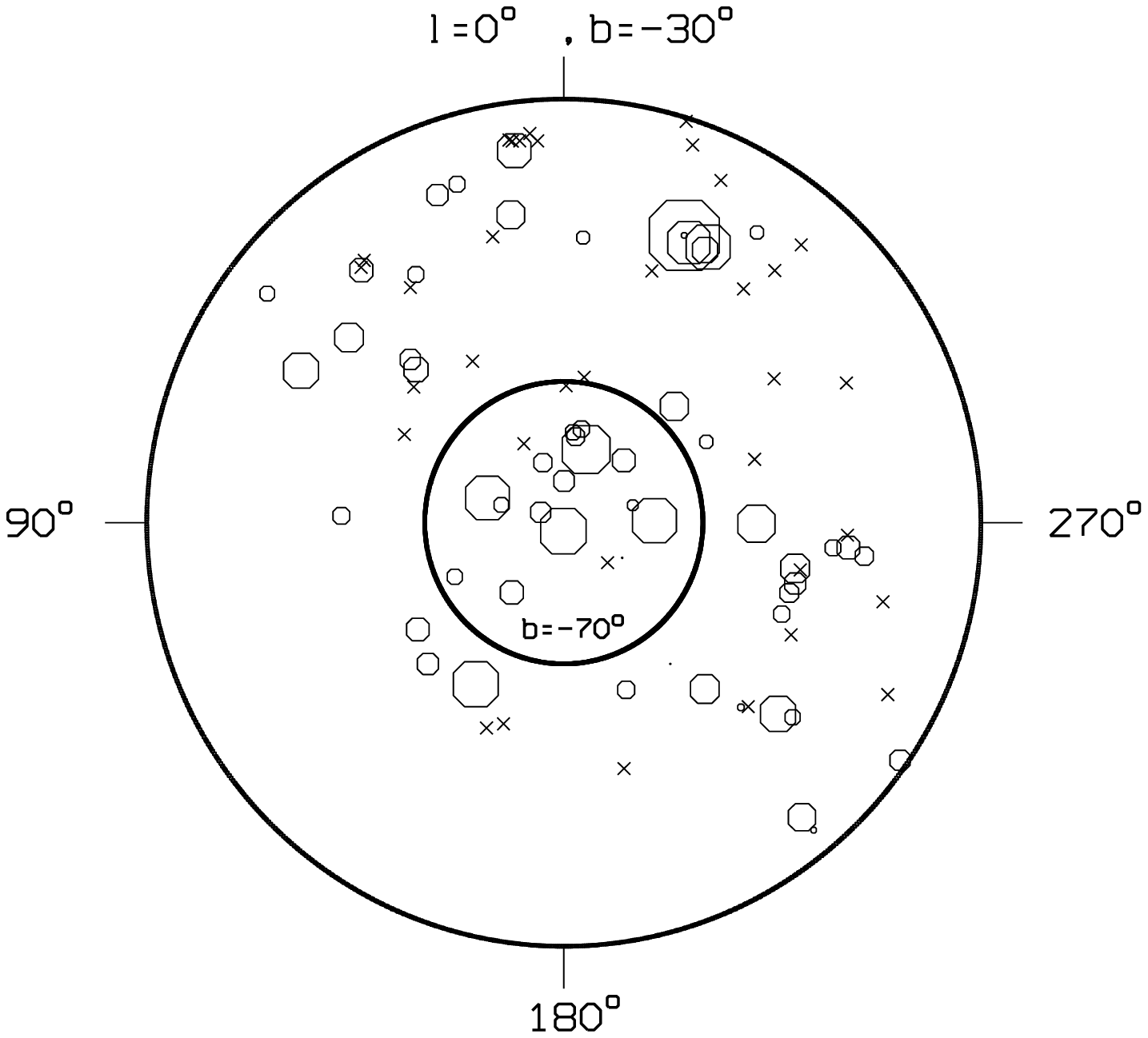,width=9.0cm,angle=270}}
\caption[]{The distribution on the sky of the directions towards the
95 clusters with b$\leq$-30$\degr$ (see paper II). For 57 clusters a 
background density is known (circles) and the other 38 clusters for 
which no background density is known are indicated by crosses. The 
size of the circles reflects the background density.}  
\end{figure}

As can be seen from Fig. 4, there is some structure in the
distribution of the background values, in the sense that large values
of $\sigma _b$ appear to cluster on scales of about 20 degrees, which
corresponds about to 100 Mpc for z=0.1. However, around the south
galactic pole, both high and low background values occur, and this is
also the case around l $\simeq 338\degr$, b $\simeq -46\degr$. The
latter concentration of high backgrounds corresponds to a supercluster
mentioned in paper I; another, at l $\simeq 255\degr$, b $\simeq -54$
corresponds to the Horologium-Reticulum supercluster (Lucey et al.,
1983), that was also mentioned in paper I.

\subsection{Which profile fits best?}

The MLM technique does not provide an absolute estimate of the
probability that a distribution of galaxies is `produced' by a certain
type of profile. However, it does allow a comparison of the relative
merits of different profiles, through the {\em likelihood ratio}
statistic (see, e.g.  Meyer 1975). If there are two alternative
hypotheses that we are testing on a single data-set, the {\em
likelihood ratio} statistic allows one to estimate the probability
that the hypothesis with the lowest likelihood is false, under the
assumption that the hypothesis with the highest likelihood is true.

In practice, one computes the statistic $-2 \ln (L_1/L_2)$ from the
values of $-2 \ln L$ given in Tables 4 to 7, for the different density
profiles, taking one of the values of the likelihood found for a given
data-set as a reference $L_1$, and comparing that to
the likelihood $L_2$ found for the same data-set and a different
profile. For large samples, the statistic $-2 \ln L_1/L_2$ has a $\chi^2$ 
distribution with $n$ degrees of freedom, where $n$ equals the number of 
parameters not restricted by the null hypothesis.

In the present case, the parameters that are fixed for all density
profiles (i.e. the centre position, the ellipticity and position
angle) do not contribute to the degrees of freedom of the
distribution, and $n=3$ because the core-radius (or characteristic 
radius) $r_c$, the slope $\beta$, and
background density $\sigma_b$ are estimated independently for the
different profiles. Note that the central density is {\em not} a free
parameter, as it is constrained by the normalization implicit in the
MLM technique.

In general, the likelihoods obtained from the fitting of the King
profile are higher than those obtained from the fitting of the other
profiles: in 37/45, 34/50 and 38/51 cases, respectively, comparing the
King profile with the Hubble, NFW and de~Vaucouleurs profiles.
However, according to the {\em likelihood ratio} statistic the
differences in the likelihood values obtained from fitting different
profiles to the same cluster data-set, are generally not statistically
significant (Fig. 5).

\begin{figure*}
\vbox
{\psfig{file=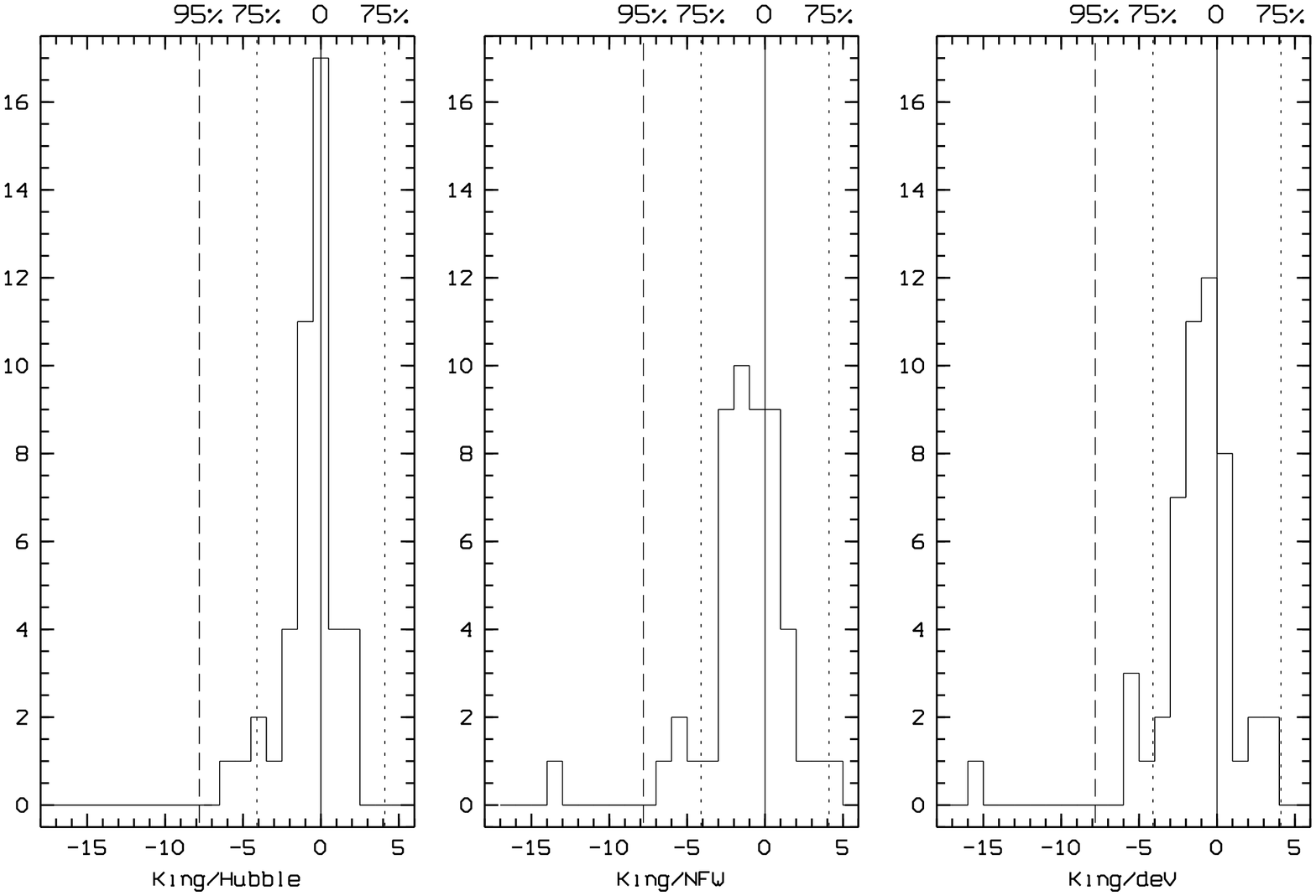,width=19.0cm,height=10.0cm,angle=270}}
\caption[]{Distributions of $-2 \ln L_1/L_2$ for the comparison between
King and Hubble profiles (left), between King and NFW profiles
(centre), and between King and de~Vaucouleurs profiles (right).
Negative values indicate that the first profile is preferred. We also
show 95 $\%$ (long dash) and 75 $\%$ (short dash) significance
levels. We note that for the right histogram, A3141 is out of the range with 
a value of -25.72.}
\end{figure*}

If we set a 95~\% significance level for rejection, and take the King
profile as our null hypothesis, we do not reject the Hubble profile as
inferior to the King profile for any of the clusters. For one cluster,
A3528, the King profile provides a significantly better fit than the
NFW profile, and for two clusters (A3128 and A3141) the
de~Vaucouleurs profile is rejected. There is, however, no
cluster for which the Hubble, the NFW or the de~Vaucouleurs profile provides a 
significant better fit than the King profile.

Most of the individual cluster density profiles, considered on the
entire available area, do not allow us to select one of the four model
profiles as the clear favorite and reject the others. However, in most
cases the likelihood of the fit is highest when the King model is
used. This may well be an indication that our inability to choose a
particular model profile is due more to limited statistics than to a
fundamental problem with the selection of a best-fitting profile.
Moreover, we know that the shapes of the four profiles are very
similar in the outer parts of the clusters, beyond 1 or 2 King core
radii. The differences between for example a King and an NFW profile
occur mainly in the very central parts of the clusters. 

We have re-computed the statistic $-2 \ln (L_1/L_2)$ inside a square
of 2 King core radii for the King and NFW profiles. In general, the
differences between the fits are more significant in the inner
regions: 32 out of 50 clusters show a better fit for the King profile
at a confidence level of 99$\%$, while only 6 out of 50 yield a better
fit for the NFW profile at the same confidence level. This again seems
to point to the existence of a core in a large fraction of clusters of
galaxies.

\section{Density profiles of composite clusters}

The result in $\S$ 3.4, viz. that the likelihood ratios for the King
and NFW profile fits, when taken at face value, all indicate that the
King profile provides a better fit than the NFW profile is quite
tantalizing, although we realize full well that by themselves none of
the likelihood ratios except one or two really indicate a highly
significant preference for the King profile over the NFW profile.
However, it cannot be excluded that the King profile in fact provides
a better fit for a majority of our clusters, but that the limited
statistical weight of the data for most of our clusters, when taken by
themselves, prevents a significant demonstration of that fact. The
fact that the King profile is preferred over the NFW profile in the
central regions is quite interesting, because this result is obtained
with smaller numbers of galaxies.

By combining the galaxy distributions for many clusters we have
increased the statistical weight, in an attempt to get a more
significant answer to the question if certain model profiles fit
better than others, and particularly if galaxy distributions in
clusters have cores or central cusps.

\subsection{The construction of the composite clusters}

The combination of the galaxy distributions of many clusters to
produce what we will refer to as a composite cluster, requires a lot
of care. In order not to smooth (or produce) possible cusps, and to
avoid artefacts as a result of differences in sampling or due to the
superposition of elongated, imperfectly centered galaxy distributions,
we have used the following procedures.

First, we must account for the fact that different clusters have
different `sizes'. If one were certain that all clusters obey the same
type of projected galaxy density profile, with different `core radii'
to be sure, one could simply put the projected distances between the
galaxies in all clusters on the same scale, by using the core radius
to scale all projected distances before adding the galaxy
distributions. However, one can also make a case for scaling by
r$_{200}$, the radius within which the average density is 200 times
the average density in the universe. According to Navarro, Frenk and
White (1997), scaling with r$_{200}$ takes into account differences in
mass. In the present discussion, we have produced composite clusters
without scaling. Only for the comparison between the composite
clusters, i.e. for getting an indication which profile best fits the
composite clusters (see $\S$ 4.4), we have applied scaling.

A second result of the differences in characteristic scale is that the
observations of the various clusters do not cover the same aperture,
when expressed in core radius. To correct for this, we have applied a
slightly modified version of the method devised by Beers and Tonry
(1986), and by Merrifield \& Kent (1989). The essence of that method
is that one adds a model contribution to the observed data in the area
where a particular cluster does not contain data, based on its
relative contribution in the area where it does contribute. In
practice, we combine the galaxy distributions of the clusters,
beginning with the cluster with the largest and second largest core
radii. We add some galaxies to the second largest cluster in the area
where this cluster does not have data, according to the density
profile that was fitted to the central region. For this
`extrapolation' we used the King-profile fit, but as all model
profiles that we used have essentially the same outer logarithmic
slope, the result would have been identical had we used e.g. the NFW
profile fit. This process is repeated for all remaining clusters in
order of decreasing core radius.

In this method there is an evident problem with the background. When
we "reconstruct" the profile as described above, the background
galaxies are implicitly, and approximately taken into account. The
reason for this is that the galaxies that are added in the outer
regions of the smaller clusters, are added in proportion to the total
number of galaxies, i.e. cluster members {\em plus} background
galaxies. This causes the background to be not exactly `conserved'.
However, it is not simple or even possible to do much better without
redshift information. To estimate the background effect, we will build
a composite cluster (see $\S$ 4.3) with only ENACS galaxies, for which
the membership is clear from the redshift.

\subsubsection{The effect of cluster elongation}

We have taken the elongation of the clusters into account by
`circularizing' the individual galaxy distributions by increasing all
projected distances orthogonal to the major axis by the appropriate
factor, thus `expanding' the distribution parallel to the minor axis.
For this correction we used the ellipticities from Tab.  1. Even
though the ellipticities of most of our clusters are not very large,
this correction may be important because the superposition of
elliptical galaxy distributions with randomly distributed orientations
will cause the outer densities to be underestimated with respect to
the inner ones, which may produce an artificial cusp (see Fig. 6).

\begin{figure} 
\vbox {\psfig{file=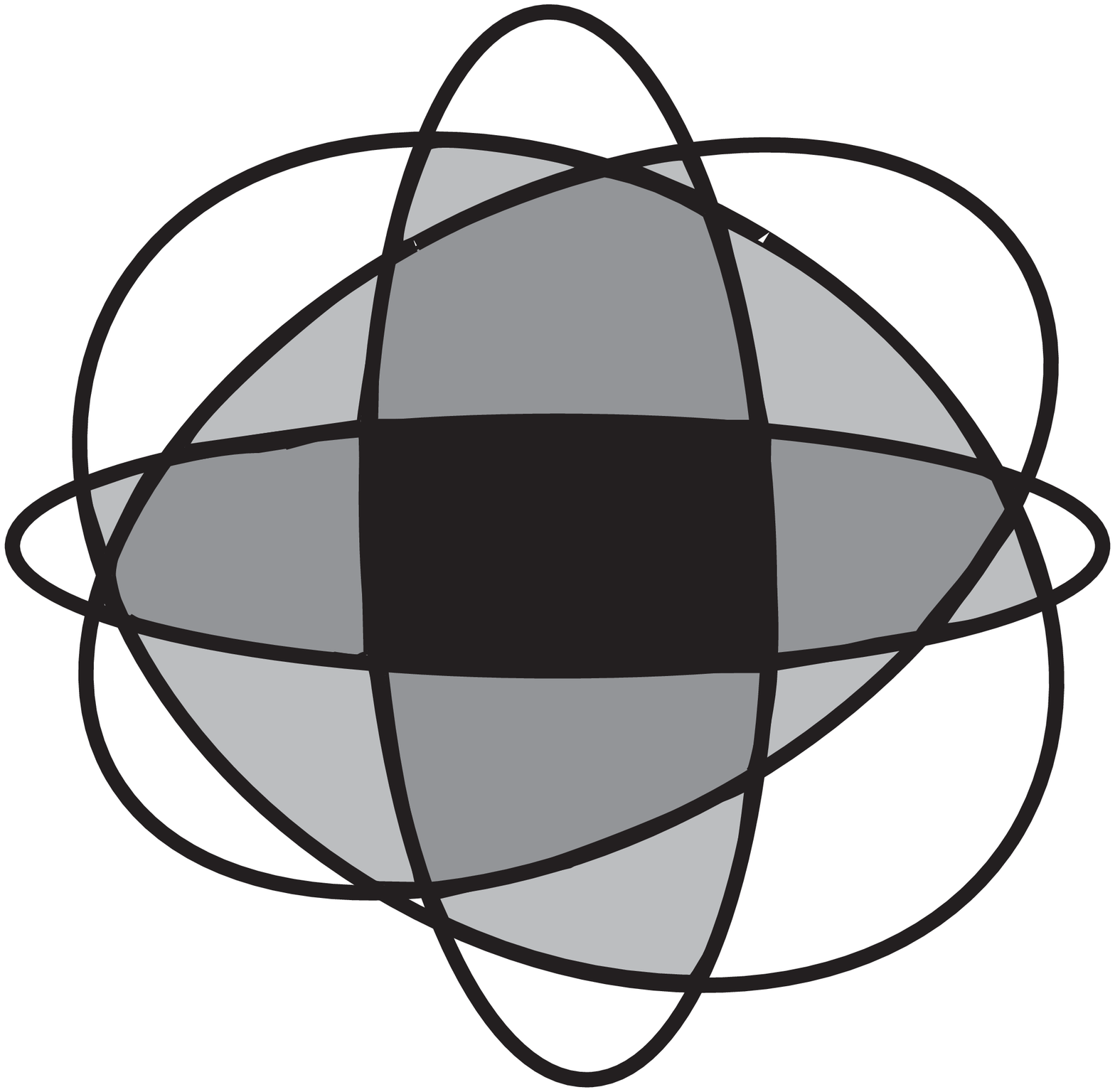,width=8.0cm,angle=200}}
\caption[]{Superposition of 4 elliptical clusters with different
orientations of the major axes. The intensity of the shading is
proportional to the number of contributing clusters.}  
\end{figure}

We have checked the importance of the ellipticity correction by
constructing two artificial composite clusters. Both are the
superposition of 50 `perfect', artificial King-profile clusters, each
containing 100 galaxies in a 1200 kpc square aperture (without
background). We assumed $\beta =1.0$ and selected r$_c$ from a uniform
distribution in the (50,150)-kpc interval. One composite cluster was
made with an ellipticity distribution which mimics the observed
one. I.e., we assumed 35 clusters to have $e$ uniformly distributed in
the interval (0.0,0.1) and the other 15 uniformly in the interval
(0.25,0.75). We assumed position angles to be randomly distributed
between 0$\degr$ and 360$\degr$, which is fully consistent with the
observations. For the first composite cluster, no correction for
ellipticity was applied, the galaxy distributions were simply summed.

For the second artificial composite cluster we also used 50 clusters,
again with $\beta =1.0$ and r$_c$ from a uniform distribution in the
(50,150)-kpc interval. However, in this case, all ellipticities were
drawn uniformly from the range (0.0,0.1). In Fig. 7, we show the
profiles of the two artificial composite clusters. As expected, the
composite cluster with the real ellipticity distribution, and without
correction for ellipticity, has a 2$\sigma$-excess in the very centre
compared to the artificial cluster built from almost round
clusters. Adding 15 elongated clusters thus induces a small but
significant cusp.

\begin{figure} 
\vbox {\psfig{file=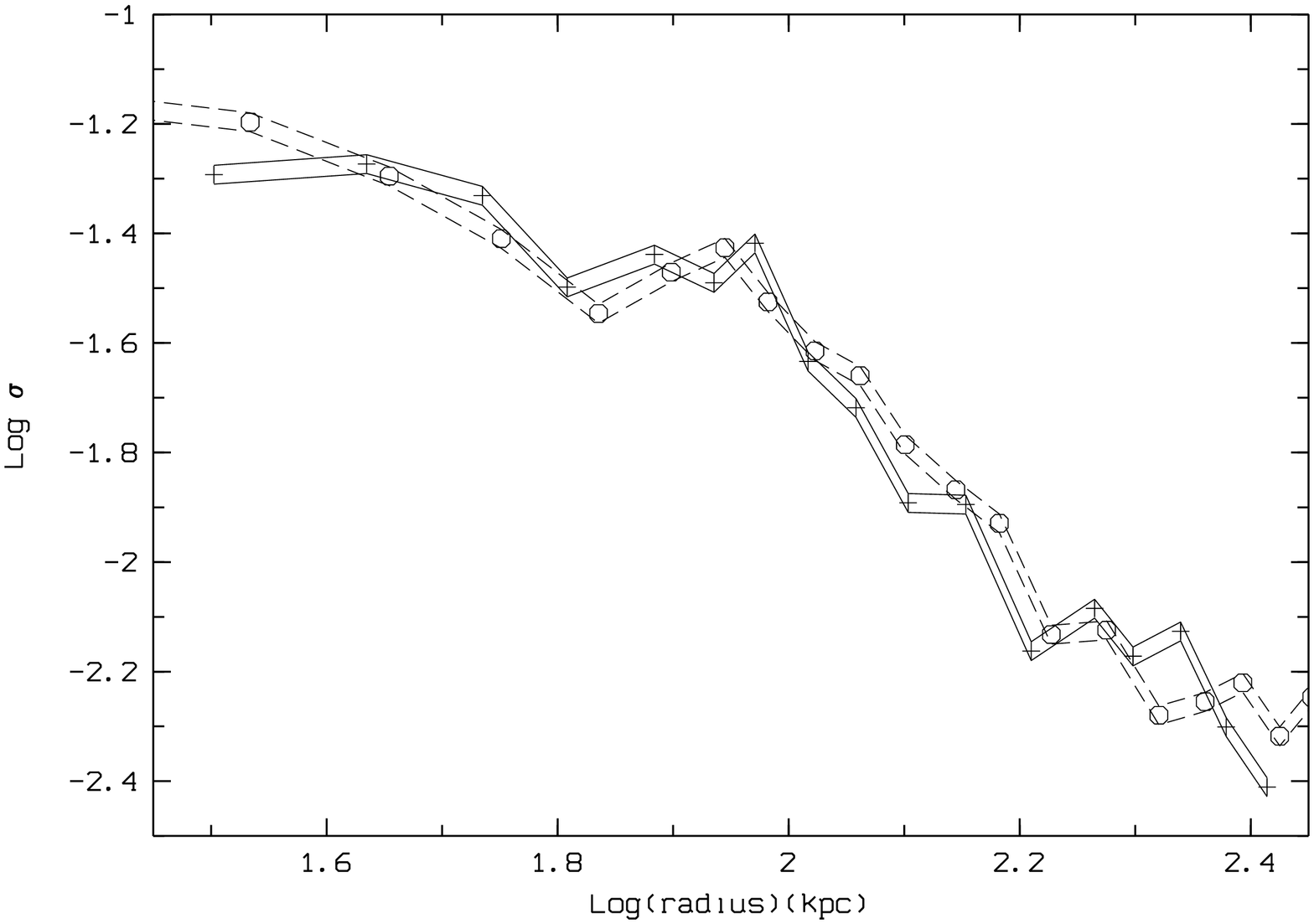,width=8.0cm,angle=270}}
\caption[]{The two artificial composite clusters, each built from 50
`perfect' King-profile clusters of 100 galaxies. The profile indicated
by circles is the result of taking the observed ellipticity
distribution, but without applying a correction for ellipticity. The
profile indicated by crosses is the result of adding 50 clusters which
are all practically round, making a correction for ellipticity
unnecessary. Errors are shown as dashed and solid lines.}
\end{figure}

\subsubsection{The effect of centering errors}

Whereas neglecting the ellipticity correction produces an artificial
cusp, errors in the central position will tend to destroy (if not
totally, at least partly) a real cusp in a composite cluster, as
argued by Beers and Tonry (1986). The magnitude of this effect depends
fairly strongly on the ratio of the position errors and the core
radii. We have tested the effect of errors in the centre position for
our dataset, using again `perfect', artificial NFW-profile clusters.
We use 50 perfectly circular clusters with 100 galaxies each. The
characteristic radii are uniformly distributed in the range (250,300) kpc. 

We produce five composite clusters, the first with perfect centering and
the other with random shifts of the real centres lower than 40 kpc, 65 kpc, 
85 kpc, 125 kpc and with arbitrary orientation. The profiles for the two first 
artificial composite clusters are shown in Fig.~8, from which we conclude that
there is indeed a smoothing due to the position errors, but it is
very small. Therefore it is very unlikely that our results are
seriously influenced by it. Actually, a 2-D Kolmogorov-Smirnov 
test indicates that the two artificial composite clusters are
indistinguishable at a level of about 99\%, but this result refers to
the whole area, and not just the central part of the clusters.
In order to quantify the effect of larger shifts, we have compared the
quality of the NFW and King profile fitting for the 5 composite clusters
through the likelihood ratio statistic (see Sec. 3.4). We find that the first 
3 clusters (shifts of 0, 40 and 65 kpc), are significantly better fitted by 
a NFW profile than by a King profile at the 99$\%$ confidence level. A shift of
85 kpc provides also a better fitting for a NFW profile but only at the 
confidence level of 95$\%$. Finally, for shifts of 125 kpc, we have a slightly 
better fitting for a King profile, even if the difference is not significant. 
This shows that even if we attribute the entire difference between our isodensity 
centres and the ROSAT centres to errors in our centre positions (which is 
certainly an overestimation), those errors are not likely to have destroyed a 
potential cusp. 

\begin{figure} \vbox
{\psfig{file=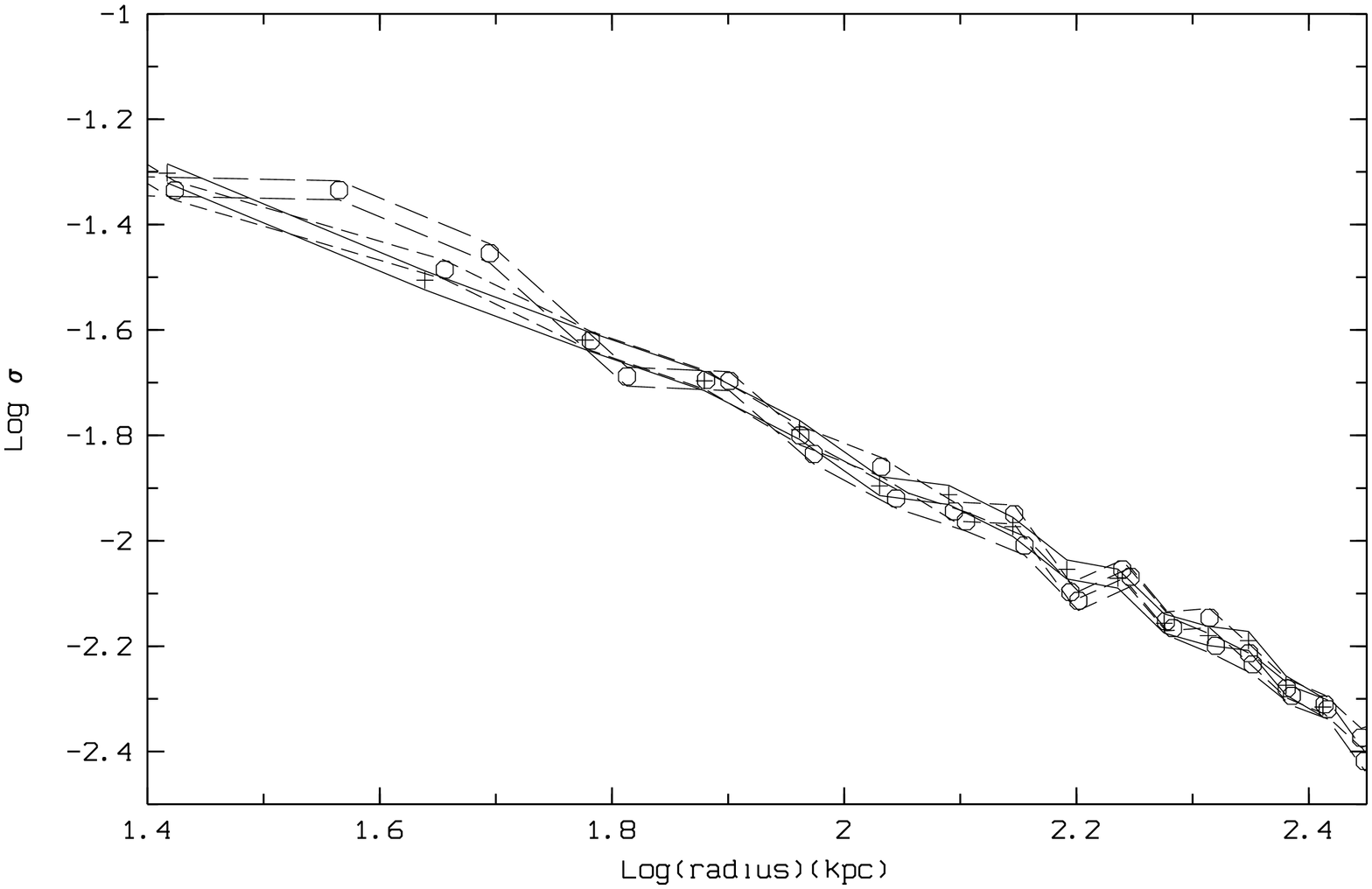,width=8.0cm,angle=270}}
\caption[]{Superposition of the profiles calculated for composite
clusters that were built from artificial clusters with NFW-profiles 
with true centres (solid line), with moderate centre shifts
(40 kpc: short-dashed line) and large centre shifts 
(125 kpc: long-dashed line).}
\end{figure}

\subsection{The COSMOS composite cluster}

For the construction of the composite cluster based on COSMOS data, we
have used the 29 clusters that we also used in our analysis of the
Fundamental Plane of clusters (Adami et al. 1998). This subset of 29
clusters was selected from the sample of 62 clusters in Tab. 1 on the
basis of the regular character of the galaxy distribution. 
For these 29 clusters, we have indicated in Fig.~1 the areas that we used
in the construction of the composite cluster. It can be seen that these are 
free from evident substructures. In order to quantify this fact, we have applied a 
Dressler-Shectman test inside these areas. At a confidence level of 5$\%$, there
are only 4 clusters among the 29 which show signs of substructures: viz. A1069, 
A2644, A3122 and A3128 (14 $\%$ of all the galaxies). If we exclude the Emission 
Line Galaxies only A1069, A2644 and A3128 show substructure. The contribution 
of these clusters is certainly minor and we considered the composite cluster 
essentially free from substructures.
A3128,
which is classified as multimodal, is included as its main peak is
well-defined. The 29 clusters all have a redshift smaller than 0.1 and
at least 10 redshifts in the area within 5 King core radii. In principle,
the latter requirement is not important for the present discussion,
but if one adds the 15 regular clusters from Tab. 1 with less than 10
redshifts, the total number of galaxies increases by only 13\%.

From the 29 clusters with 4735 galaxies, we have constructed a
composite cluster; in doing so we added 2505 galaxies in the outer
parts of the smaller clusters (see $\S$ 4.1), to produce a cluster
with 7240 COSMOS galaxies. The area where we fit the models on the composite 
cluster has a radius of 600 kpc, which is similar to that of the area over
which the profiles of the individual clusters were fitted. 
For the present discussion no scaling of projected distances was
performed. Although the dispersion of the characteristic scales is not
very large (see $\S$ 3.2), the superposition of galaxy distributions
with different scales may affect the profile of the composite cluster.
For example, adding clusters with identical types of profile with
small and large core radii might produce a cluster which does not have
the same profile as the individual clusters from which is was built.
When discussing the question of which type of profile best fits the
observations (see $\S$ 4.4) we will therefore also discuss results for
composite clusters for which the constituting clusters were scaled
before adding them.

In Fig.~9 we show the result for the COSMOS composite cluster, for
which all projected distances were scaled with the King and NFW 
characteristic radii of the individual clusters: we take the value predicted
with the regression line between the King and NFW radii. We note here that the
relation between these two radii is well defined (see $\S$ 3.2). The profile 
was calculated in radial bins which
each contain 20 galaxies. The dashed lines indicate the 1-$\sigma$
range around the observed values (the dots). The two full-drawn curves
represent the best King and an NFW fits. It is
clear that the observed values within r$_c$ (log r $<$ 0) indicate a
flat profile. In $\S$ 4.4 we will discuss the result of a formal test
of which of the two model profile best represents the three composite
clusters. Looking at this figure, one may already get some idea about
the outcome of that test.

\begin{figure} 
\vbox
{\psfig{file=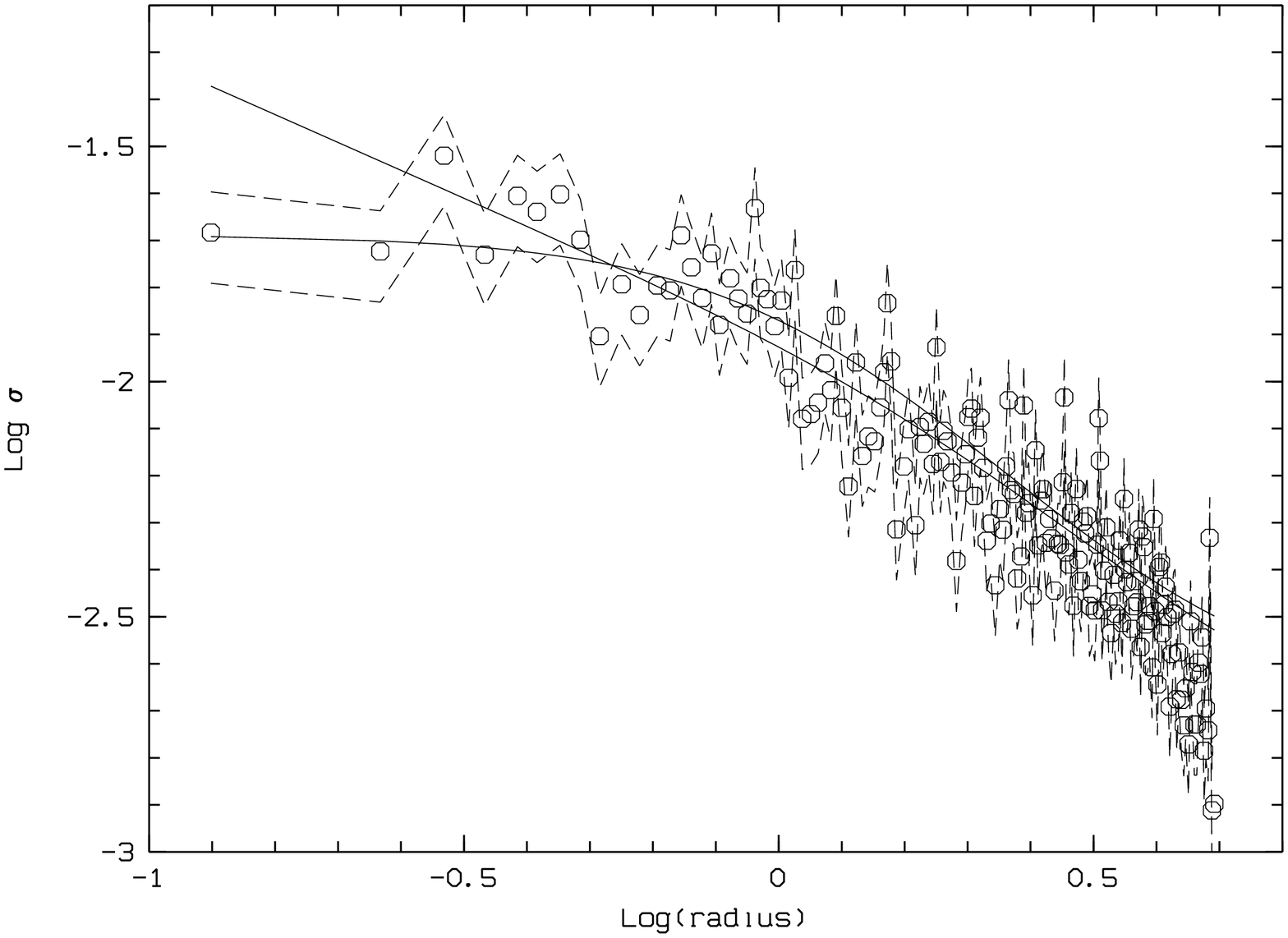,width=8.0cm,angle=270}}
\caption[]{The projected density in the composite COSMOS cluster that was
produced by scaling all projected distances with the core radii of the
individual clusters, so the horizontal scale gives r/r$_c$. The dots
give the observed values in bins which contain 20 galaxies each. The
dashed lines indicate the 1-$\sigma$ interval on either side of the
observed values. The full-drawn curves represent the King and an NFW
profile.}
\end{figure}

Using the MLM method we fitted both a King and an NFW profile to the
(unscaled) composite COSMOS cluster. These fits give: $\beta =1.00\pm
0.02$, r$_c=89\pm 5$ and $\sigma _b=2.00\pm 0.03$ 10$^{-3}$ for the
King profile, and $\beta =0.56\pm 0.01$, r$_c=318\pm 34$ and $\sigma
_b=1.44\pm 0.05$ 10$^{-3}$ for the NFW profile. The value of $\beta$
of 0.56$\pm$0.01 for the NFW-profile is somewhat lower than the value
one would predict on the basis of the $\beta$ for the King-profile
fit, using the expected relation between the two $\beta$'s. Imposing 
$\beta$ = 0.67 for the NFW profile, the fit is worse but we
obtain $\sigma_b=1.70$ 10$^{-3}$, which is closer to the value found
in the King-profile fit. This shows that $\beta$ and $\sigma _b$ are
correlated.

The values of r$_c$, $\beta$ and $\sigma_b$ for the composite cluster
are globally consistent with the average values found for the 62
clusters in $\S$ 3.2. The agreement is also good if we compare with
the average values for the subsample of the 29 clusters used here
rather than for the total sample. For the 29 clusters, the average
values of $\beta$ and r$_c$ and the value of $\Sigma \sigma{_b{_i}}$
are 1.04$\pm$0.08, 115$\pm$67 kpc, 1.17$\pm$0.56 10$^{-3}$ for the
King-profile fits and 0.61$\pm$0.05, 276$\pm$181 kpc, 1.13$\pm$0.55
10$^{-3}$ for the NFW-profile fits. 

If we restrict the fits to the central region (radius: 300 kpc) of the
composite cluster (924 galaxies), we obtain $\beta$ = 1.02 and r$_c$ =
106 kpc for the King profile and $\beta$ = 0.55, r$_c$ = 272 kpc for the NFW
profile. These values for the central region agree better with the
values for the individual clusters; this may be partly due to the fact
that the addition of galaxies in the outer regions of the clusters
hardly affect the central region. But the relative importance of
errors in the assumed centre positions or in the core radii
determinations is probably larger.

We have also made a composite cluster frçom those 14 clusters for which
a central position is available from ROSAT, using the latter centre rather 
than ours. We fit both a King and a NFW profile in a more central area 
(radius: 200 kpc) and find a better fit for the King profile at the 95$\%$ 
confidence level. The use of the ROSAT centres does not change our 
conclusion about which profile fits best.

\subsection{The ENACS composite cluster}

In principle, the ENACS data could have been used to make independent
solutions for r$_c$ and $\beta$ for the individual clusters. Because
the ENACS data allow us to select cluster members through their
redshifts, the backgrounds would, by definition, be zero. However, the
ENACS data are limited to the central regions of the clusters, and
redshifts are not available down to the magnitude limit of the COSMOS
data. Therefore, the number of ENACS galaxies in a cluster is, in
general, too small to make a reliable fit. By combining the ENACS data
for all 21 clusters among the 29 for which the ENACS data cover a
rectangular area of at least 400 kpc to a side we have constructed an 
ENACS composite cluster with 388 galaxies.

MLM fits to this composite cluster yield $\beta = 1.00\pm 0.05$ and
r$_c=91\pm 12$ for the King profile and $\beta =0.51\pm 0.03$,
r$_c=274\pm 61$ for the NFW profile. When we impose $\beta$ = 0.67 for
the NFW-profile fit leads to a value of the likelihood that is a
factor 5 worse than the maximum likelihood. It is likely that the
lower value of $\beta$ is, at least partly, due to the fact that the
size of the aperture is fairly small compared to the value of r$_c$
for the NFW-profile fit.

Although the statistical weight of the ENACS composite cluster is much
less than that of the COSMOS composite cluster, the results of the
2-parameter, rather than 3-parameter, fit provide strong support for
the profile parameters found in $\S$ 4.2.

\subsection{Which profile best describes the observations?}

Using the COSMOS composite cluster, we have again addressed the
question posed in $\S$ 3.4. We use three different versions of the
composite cluster: one without scaling (as in $\S$ 4.2), one in which
all projected distance are scaled with the values of r$_c$ for the
individual clusters as derived in $\S$ 3.2 (and given in Tabs. 4 to
7), and one in which we scaled with the individual values of
r$_{200}$. The latter were calculated from the ENACS velocity
dispersions, following e.g. Carlberg et al. (1997). For all three
composite clusters we have made fits with a King- and an NFW-profile
model.

We thus derived three values for the likelihood ratio $-2 \ln
(L_1/L_2)$. These are -17, -17 and -10 for unscaled, r$_c$-scaled and
r$_{200}$-scaled composite clusters, respectively. All three values
are negative, and thus indicate a preference for the King profile over
the NFW profile. The formal associated significance levels are 99, 99
and 95 percent. So, at face value, the result of $\S$ 3.4 is confirmed
and amplified: that the majority of our clusters are indeed better
explained by King profiles than by NFW profiles. 

The following caveats must, however, be remembered in connection with
this conclusion.

First: there is a small smoothing effect due to the errors in the
centre positions, which we cannot quantify very well (using the best-fit 
centres is the best that we can do). Secondly, our
selection of clusters is certainly not unbiased: we have used the most
regular clusters (which in general are among the most massive
ones). This could produce a sample of well-evolved clusters, which
need not be typical of the total rich cluster population as far as the
galaxy density profile is concerned.  Thirdly, the assumption
underlying the composite clusters, namely that when properly scaled 
, all clusters have the same profile, may well be a simplification.
Finally, the likelihood ratio for the two fits to the composite
clusters with r$_{200}$-scaling is least indicative of a preference
for a King profile over an NFW profile. This may be just a statistical
effect, but it might also be an indication that the absence of
scaling, or the r$_c$-scaling, cause some smoothing due to incorrect
combination of profiles which, intrinsically, have a moderate
cusp. The likelihood ratio for the r$_{200}$-scaling applies to a
composite of the 7 clusters in which the data extend out to at least
0.75 r$_{200}$ (this cluster contains 1092 COSMOS galaxies).

\subsection{The effect of absolute magnitude}

Finally we calculate likelihood ratios for different ranges of
absolute magnitude. The reason for this is that e.g.  Carlberg et
al. (1997) claim the existence of a central cusp for the bright
galaxies in clusters. We use the COSMOS composite cluster and we fit
King and NFW profiles in the central 300 kpc (radius), for different magnitude
ranges. We define the following 4 intervals of absolute
magnitude, which contain 200 galaxies each: (-22.4,-18.78),
(-18.77,-18.03), (-18.02,-17.48) and (-17.47,-16.89). The likelihood
ratios in these intervals are, +1, --8, --8 and --4, respectively. It
is clear that the preference for a King profile is certainly not
shared by the brightest galaxies.

We have tried to estimate the magnitude for which the galaxy profile
seems to change character. Taking narrower intervals, which contain 50
galaxies each, we find that for absolute magnitudes brighter than
-18.4$\pm$0.2 the likelihood ratio (or rather $-2 \ln (L_1/L_2)$) is
about 0, while fainter than -18.4$\pm$0.2 we obtain values between -4 and -8.
That bright and faint galaxies do not have the same distribution is
not totally new. E.g., Biviano et al. (1996) and Dantas et al. (1997)
find that the faint and the bright galaxies in the Coma cluster and in
A3558 respectively, have different distributions. The bright galaxies
are very concentrated around a few specific positions while the
distribution of the faint ones is smooth, and without a cusp.

\subsection{Comparison with results in the literature}

From the discussions in $\S$ 3.4 and $\S$ 4.4 we conclude that the
COSMOS and ENACS data are better fitted by a King profile (or more
generally: a profile with a core) than by an NFW profile (or rather: a
profile with a cusp). In other words: our data seem to suggest a
logarithmic slope in the central cluster region that is flatter than
--1. Admittedly, this conclusion is based primarily on the composite
clusters, and therefore we cannot be sure that the conclusion is true
for all of the individual clusters as well. On the other hand, the
results of the discussion in $\S$ 3.4 suggest that the conclusion may
well be valid for most of the individual clusters. In
addition, there is some evidence that the intrinsically brighter
galaxies have a more cusped profile than the fainter ones.

In the literature there are several results that seem to be at odds
with our first conclusion. Beers and Tonry (1986) argue e.g. for the
presence of cusps and Carlberg et al. (1997) concluded that the CNOC
clusters at a redshift of $\approx$ 0.3 have galaxy density profiles
that are quite consistent with the NFW profile. Both conclusions are 
based on composite
clusters, and as we discussed in $\S$ 4.1 several details of the
construction of a composite cluster influence the destruction of cusps
or their artificial production. Neither Beers and Tonry, nor Carlberg
et al. seem to have taken the elongation of the clusters into account.
As we found in $\S$ 4.1.1, that may be (partly) responsible for the
appearance of a small cusp. On the other hand, it is also possible (if
not likely) that the differences are largely due to the fact that our
galaxy samples extend to fainter absolute magnitudes. We note here
that the limiting magnitude of the CNOC survey (Carlberg et al., 1997)
matches well with the value given in $\S$ 4.5 for which the King and NFW
profiles become equivalent.

Therefore, the apparent disagreement may be largely resolved by our
second conclusion, viz. that the brighter galaxies have a more cusped
distribution than the fainter ones. Incidentally, this latter
conclusion would also seem to indicate that for our clusters, the
errors in the centre positions are indeed sufficiently small that they
do not erase the cusp in the distribution of the brighter galaxies.

The values that we find for r$_c$ are consistent with the results
obtained by Girardi et al. (1995) and Bahcall (1975). Our value for
$\beta$ is only moderately consistent with that obtained by Lubin \&
Bahcall (1994), who found 0.8$\pm$0.1. This discrepancy may be due to
the fact that these authors fit their profile in the outer regions
(500 to 1500kpc), which may contain substructure.

Our values of r$_c$ and $\beta$ are local (z $<$ 0.1), and our $\beta$
value can be compared with $\beta$ estimates for other epochs.
Recently, Lubin and Postman (1996) have studied 79 distant clusters
from the Palomar Distant Cluster Survey (PDCS) with estimated
redshifts between 0.2 and 1.2, where the precision of the redshift
estimation was 0.2. They found r$_c=50$ kpc and $\beta =0.7$ using a
King profile. Because they have estimated the background at 1.2 Mpc
from the centre, their estimate of r$_c$ is probably not very
accurate, and we do not consider the difference between their and our
value for r$_c$ to be really significant.

If one accepts the formal error bars for $\beta $, the difference in
the $\beta $ values for distant and nearby clusters would be
significant. If the differences between the values of $\beta$ are
real, this indicates a change of the outer slope of the density
profile with redshift (Tab. 8), in a way that is consistent with
global ideas about an increase of the concentration of the clusters
with cosmological time. According to commonly adopted formation and
accretion scenarios for clusters of galaxies, young clusters have a
flatter density profile than older, more evolved clusters.

\begin{table} 
\caption[]{$\beta$ values for different redshifts from the present
study and the study of Lubin \& Postman (1996).}  
\begin{flushleft}
\small 
\begin{tabular}{crrrr} 
\hline 
\noalign{\smallskip} 
redshift & 0.07$\pm $0.05 & 0.3$\pm $0.1 & 0.6$\pm $0.1 & 1$\pm $0.2 \\ 
$\beta $ & 1.02$\pm $0.08 & 0.73$\pm $0.17 & 0.74$\pm $0.17 & 0.68$\pm $ 
0.40 \\ 
\hline 
\normalsize 
\end{tabular} 
\end{flushleft}
\end{table}

\section{Profile slope $\beta$ and $\Omega_0$}

From the analysis in $\S$ 3.2, $\S$ 4.2 and $\S$ 4.3 it is clear that
the outer slopes of the galaxy distributions of the individual (and
composite) clusters are all very much consistent with a value of 1.0,
for King- and Hubble-profile fits. In addition, there is satisfactory
consistency with the $\beta$-values for NFW- and
de~Vaucouleurs-profile fits (certainly for the individual clusters).
We therefore conclude that this result is quite robust.

Quite a few numerical simulations have appeared in the literature from
which the average present-day slope of the density profiles that is
predicted as a function of the initial fluctuation spectrum, the
cosmological parameters etc., can be estimated. In the following, we
compare our result for $\beta$ with predictions obtained by Walter and
Klypin (1996), Jing et al. (1995), Crone et al. (1994) and by Navarro,
Frenk and White (1995).

\begin{figure*}
\vbox
{\psfig{file=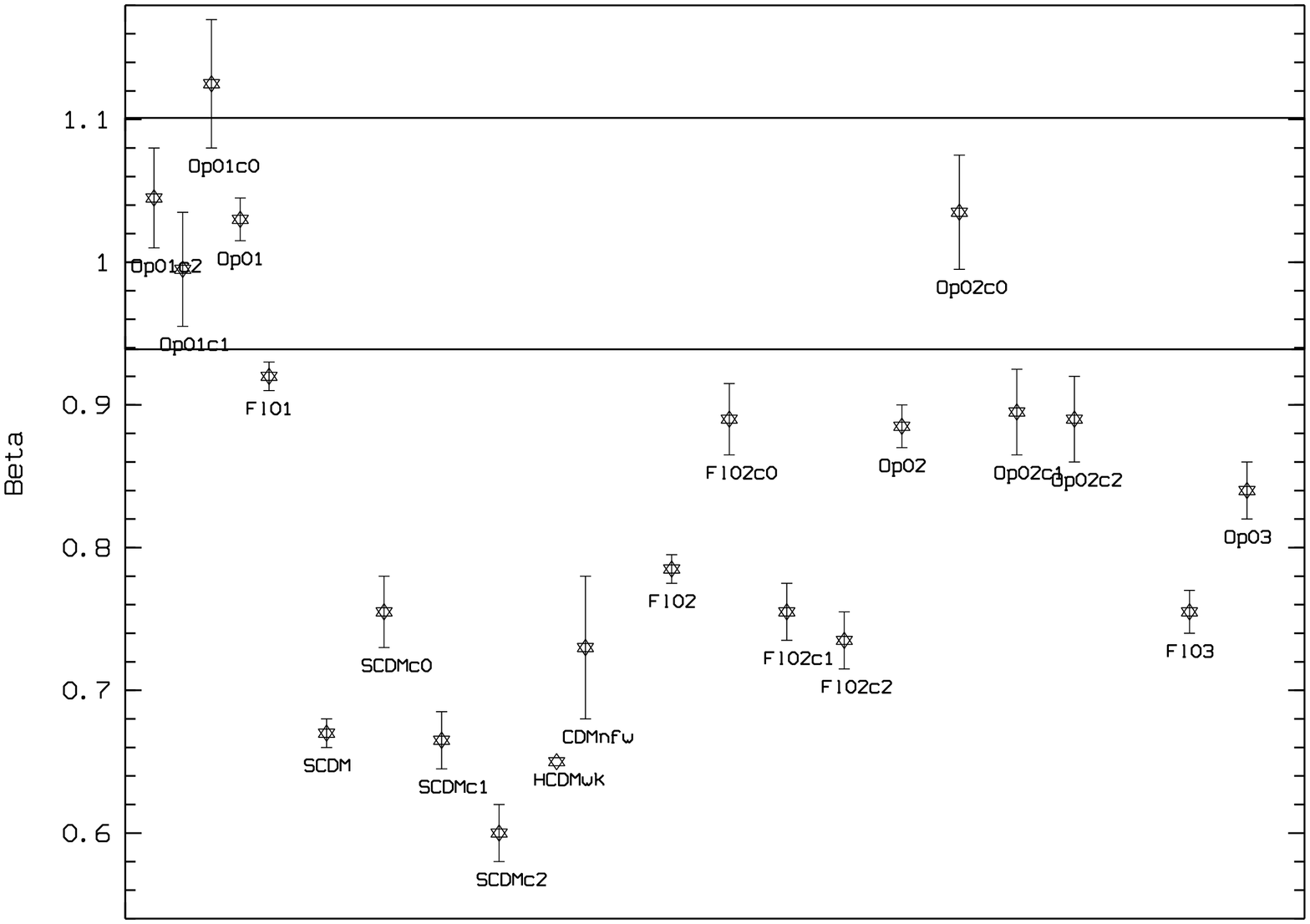,width=16.0cm,angle=270}}
\caption[]{Numerical model values and observed values for 
$\beta$. The 2 
straight lines define the range permitted with the 
invividual clusters.}
\end{figure*}

\noindent Walter and Klypin (1996) simulate a mixed HCDM model with
$\Omega _{CDM} = 0.6$ and $\Omega _\nu = 0.3$. The size of the
simulation box is between 400 and 500 Mpc and they use the
Harrison-Zeldovit'ch fluctuation spectrum. We will refer to it as the
HCDMwk model.

\noindent Navarro, Frenk and White (1995, 1996) simulate a CDM model
with $\Omega$ = 1.0. The size of the simulation box is 180 Mpc. We will 
refer to it as the CDMnfw model.

\noindent Jing et al. (1995) simulate seven scenarios, in 128 Mpc
boxes. Each model gives more than 50 clusters which is very
similar to our real sample.  The fluctuation spectrum is the
Harrison-Zeldovit'ch one. The models are: a standard CDM model with
$\Omega _0 = 1$ (SCDM), three flat models with $\Omega _0 = 0.1$, 0.2
and 0.3 (with $\Lambda $ = 0.9, 0.8 and 0.7, respectively) (Fl01, Fl02
and Fl03) and three open models with $\Omega _0= 0.1 $, $0.2$ and 0.3,
and $\Lambda $ = 0 (Op01, Op02 and Op03).

\noindent Crone et al. (1994) simulate three models with $\Omega _0 =
0.1$ and 0.2 using different fluctuation spectra: P(k)$\propto $k$^0$,
k and k$^2 $. The models are three standard CDM ones (SCDMc0, SCDMc1
and SCDMc2), three flat universes with $\Omega _0 = 0.2$ and $\Lambda
$ = 0.8 (Fl02c0, Fl02c1 and Fl02c2), three open universes with $\Omega
_0 = 0.1$ (Op01c0, Op01c1 and Op01c2) and three open universes with
$\Omega _0 = 0.2$ (Op02c0, Op02c1 and Op02c2).

All four groups fit 3-D power laws r$^{-2\beta _{3D}}$ to the outer
regions of the clusters, at the present-epoch which corresponds very
well to the redshift range of our clusters. Their values are
transformed to 2-D values, as follows: $\beta _{2D}=\frac{2\beta
_{3D}-1}2$. We compare these 2-D model values with our observed mean
$\beta _{2D}=1.02\pm 0.08$ for the King-profile fit in Fig.~10.

We find that only fairly low values of $\Omega _0$ produce average
values for $\beta _{2D}$ that are similar to the observed value; this
is in agreement with other recent results on $\Omega _0$.  However,
not all low-$\Omega _0$ models produce the observed value of $\beta
_{2D}$. Taking the results in Fig.~10 at face value, it would seem
that not only needs $\Omega _0$ to be small (a few tenths), but it is
also not likely that we need to invoke a flat Universe (i.e. a
non-zero value for $\Lambda $). The only exception to this general
conclusion seems to be the flat model with $\Omega _0$ = 0.2 and a
fluctuation spectrum P(k)$\propto $k$^0$.

It should be realized that the conclusion that $\beta_{2D}$ is best
predicted by low-$\Omega _0$ models (and possibly mostly open models
) is based on a rather incomplete range of models.  In other
words: we cannot exclude models that have not yet been calculated or
published. E.g. it is too early to firmly exclude HCDM models with
other fluctuation spectra than the one calculated by Walter and
Klypin. On the other hand, the data in Fig.~10 seem to suggest that for
$\beta$ the influence of the fluctuation spectrum is quite limited, so
our conclusion of a low-$\Omega _0$ model (most likely open) is
probably quite robust.

This value of $\Omega _0$ from the outer slope of the projected galaxy
distribution can be compared to direct estimates of $\Omega _0$ based
on cluster masses and luminosities and the field luminosity density.
The idea is quite simple: one estimates the cosmological volume 
which contains the same mass as that contained in the cluster, from a
comparison of the luminosity density in the field and the total
luminosity of the galaxies in the cluster. This yields the average
mass density which yields $\Omega _0$ through division by the critical
density $\rho_c$. In practice, the method amounts to calculating the
ratio of the $\frac ML$ ratio in cluster to the critical $\frac ML$
ratio in the field, and this result does not depend on $H_0$. A
recent application of the method was discussed by Carlberg et al.
(1996) for their sample of 16 clusters at a redshift of $\approx$ 0.3.

We have determined luminosities and velocity dispersions for our
sample of 29 ENACS clusters discussed in paper IV. The luminosities
and dispersions are calculated in an aperture of five core radii. We
derived the projected virial masses (e.g. Perea et al. 1990) and
obtained the $\frac ML$ ratios for the 29 clusters. The values of
$\frac M{L_{b_j}}$ of the 29 clusters range from about 100 to 1000;
the average value is 454, the median is 390 and the robust
bi-weight estimate of the mean is 420, in good agreement with the
value given by e.g. Bahcall et al. (1995). The local critical $\frac
ML$ ratio in the field was determined by e.g. Efstathiou et al. (1988)
and Loveday et al. (1992), who find a best estimate of 1500
$\frac{M_{\odot }}{ L_{\odot b_j}}$, with the value most probably in
the range 1100 to 2200. This yields an estimate of $\Omega _0$ on the
basis of our sample of 29 local ENACS clusters of 0.28 $\pm$ 0.19.
This value is thus quite consistent with the low $\Omega _0$ value
that we obtained from the outer slopes of the projected galaxy
distributions.

The uncertainty in the `dynamical' $\Omega _0$ estimate is to a large
extent due to the large spread in $\frac ML$ ratios for our clusters.
It appears that the individual values of $\frac ML$ correlate
moderately well with velocity dispersion, in accordance with the
result in paper IV. Expressed in $\Omega _0$ we find: $\Omega _0$ =
(3.9$\pm$1.1) 10$^{-2}$ $\sigma _v/100$ + (8$\pm$157)10$^{-3}$. If we
use the relation between luminosity, scale factor and velocity
dispersion that we derived in paper IV we obtain: $\Omega _0$ =
(3.2$\pm$0.5) 10$^{-4}$ $\sigma _v ^{1.09}$ $R^{-0.19}$, which is
totally consistent. It thus appears that there is a significant
dependence on cluster velocity dispersion in the determination of
$\Omega _0$ from cluster $\frac ML$ ratios, which could easily produce
a serious bias towards high values.

Of course, we have assumed in this analysis that in clusters light traces 
mass. Actually, this assumption is not really demonstrated, but is commonly 
used, for example in Carlberg et al. (1996). One might wonder how this 
assumption could affect our last $\Omega _0$ estimate. However, for 
the low $\sigma _v$ clusters, the present estimate is in good agreement with 
that based on the slope of the density profiles, and that gives some confidence 
that a possible bias is probably quite small.

\section{Conclusions}

We have studied the projected galaxy distributions in 77 clusters from
the ESO Nearby Abell Cluster Survey. The present sample is an unbiased
subset of the volume-limited ENACS sample, and thus forms a
representative local (z $<$ 0.1) sample of rich (R$_{ACO}$ $>$ 1),
optically selected clusters. We used both COSMOS and ENACS data to
test the character of the projected galaxy distributions. In
particular, we have investigated whether the galaxy distributions in
rich clusters have cusps or cores in their central regions.

We have made maximum Likelihood fits to the observed distribution of
COSMOS galaxies to solve for the position and the elongation of the
clusters. For 15 of the 77 clusters, no reliable centre could be
determined and these clusters were not considered further. Using the
positions and elongations, we subsequently solved for each of the 62
remaining clusters the three parameters that describe each of the four
theoretical profiles that we tested, as well as the density of
background galaxies. The four model profiles that we tested against
the data are the King, Hubble, NFW (Navarro, Frenk and White) and the
de~Vaucouleurs profiles. Although the solutions do not converge for
all of the clusters nor for all four profiles, we obtain reliable
results for between 75 and 95 \% of the clusters (depending on the
model profile).

We find mean values for r$_c$, the characteristic scale of the 2-D
galaxy distribution, and dispersions around the means of 128 $\pm$ 88,
189 $\pm$ 116, 292 $\pm$ 191 and 1582 $\pm$ 771 kpc, for the King,
Hubble, NFW and de Vaucouleurs profiles respectively. The outer
logarithmic slopes of the distributions were generalized by the usual
$\beta$-parameter, which we find to have the following average values:
1.02 $\pm$ 0.08, 1.03 $\pm$ 0.07, 0.61 $\pm$ 0.05 and 7.6 $\pm$ 0.5,
for the King, Hubble, NFW and de Vaucouleurs profiles respectively,
which are consistent. The average background density at the limit of
the COSMOS catalogue is about 4 10$^{-5}$ galaxies arcsec$^{-2}$.

In order to investigate whether the galaxy distributions in our
clusters preponderantly have cores or cusps, we have determined the
likelihood ratio for the King and NFW profiles. Using all galaxies
down to the COSMOS magnitude limit of about m$_{b_j}$ $\approx$ 19.5,
we find that in general the King profile is more likely to be a good
representation of the data than the NFW profile. However, for the
individual clusters this preference for the King profile is generally
not statistically significant. If we restrict the analysis to the
central regions, the significance of the preference for the
King-profile fits increases, even though the number of galaxies
decreases.

We have increased the statistical weight for the likelihood analysis
by combining the galaxy distributions in a subset of 29 of the 62
clusters, which show a regular galaxy distribution. We take special
care to avoid the creation of an artificial cusp (by taking the
ellipticities into account), and to avoid the destruction of a real
cusp by summing distributions with different scale lengths. We have
also checked that it is unlikely that the uncertainty in the centre
positions has erased a cusp. For the test we summed without scaling
projected distances, after scaling with r$_c$, as well as with
r$_{200}$. In all three cases we find that the King profile provides a
better fit to the data than the NFW profile, at confidence levels of
more than 95 \%. Interestingly, this preference is not shared by the
brighter galaxies.

Finally, we have used the outer profile slope (i.e. the result that
$\beta$ is very close to 1.0), in combination with several results
from numerical models to conclude that the density parameter $\Omega
_0$ is likely to be considerably smaller than unity. In addition, the
available models indicate that the Universe probably has an open
geometry (i.e. no closure through $\Lambda $ is indicated). This low
implied value of $\Omega _0$ is fully consistent with a direct
determination based on the $\frac ML$ ratios of our clusters. 

\begin{acknowledgements}

{The authors thank Harvey MacGillivray, Hans Bohringer and Doris Neumann for
providing data. AM, AB, CA and PK acknowledge financial contributions
from the French PNC, from INSU, from IGRAP, from Leiden Observatory and from
Marseilles-Provence University.}

\end{acknowledgements}

\vfill


\begin{thebibliography}{}

\bibitem{} Adami C., Mazure A., Biviano A., Katgert P., Rhee G., 1998, 
A\&A 331, 493: paper IV 

\bibitem{} Arnouts S., de Lapparent V., Mathez G., et al., 1997,
A\&AS 124, 163

\bibitem{} Bahcall N.A., Lubin L.M., Dorman V., 1995, ApJ 447, L8

\bibitem{} Bahcall N.A., 1975, ApJ 198, 249

\bibitem{} Beers T.C., Flynn K., Gebhardt K., 1990, AJ 100, 32

\bibitem{} Beers T.C., Tonry J.L., 1986, ApJ 300, 557

\bibitem{} Bellanger C., de Lapparent V., Arnouts S., et al., 1995,
A\&AS 110, 159

\bibitem{} Biviano A., Durret F., Gerbal D., et al., 1996, A\&A 311, 95

\bibitem{} Biviano A., Katgert P., Mazure A., et al., 1997, A\&A 321,
84: paper III

\bibitem{} Carlberg R.G., Yee H.K.C., Ellingson E., et al., 1996, ApJ
462, 32

\bibitem{} Carlberg R.G., Yee H.K.C., Ellingson E., et al., 1997, ApJ
485, L13

\bibitem{} Colless M., 1989, MNRAS 237, 799

\bibitem{} Colless M., Hewett P., 1987, MNRAS 224, 453

\bibitem{} Crampton D., Le Fevre O., Lilly S.J. et al., 1995, ApJ 455,
96

\bibitem{} Crone M.M., Evrard A.E., Richstone D.O., 1994, ApJ 434, 402

\bibitem{} Dantas C.C., de Carvalho R.R., Capelato H.V., Mazure A.,
1997, ApJ 485, 447

\bibitem{} den Hartog R., Katgert P., 1996, MNRAS 279, 349

\bibitem{} de Theije P.A.M., Katgert P., 1998, A\&A submitted: paper VI

\bibitem{} de Theije P.A.M., Katgert P., van Kampen E., 1995, MNRAS 273, 30

\bibitem{} Efstathiou G., Ellis R.S., Peterson B.A., 1988, MNRAS 232,
431

\bibitem{} Fletcher R., 1970, Comput. J. 13, 317

\bibitem{} Girardi M., Biviano A., Giuricin G., Mardirossian F.,
Mezzetti  M., 1995, ApJ 438, 527

\bibitem{} Hernquist L., 1990, ApJ 356, 359

\bibitem{} Heydon-Dumbleton N.H., Collins C.A., MacGillivray H.T.,
1989, MNRAS 238, 379

\bibitem{} Hughes J.P., 1997, To appear in "A New Vision of an Old Cluster: 
Untangling Coma Berenices" astro-ph 9709272

\bibitem{} Jing Y.P., Mo H.J., Borner G., Fang L.Z., 1995, MNRAS 276,
417

\bibitem{} Katgert P., Mazure A., den Hartog R., et al., 1998, A\&A
accepted: paper V

\bibitem{} Katgert P., Mazure A., Perea J., et al., 1996, A\&A 310, 8
: paper I

\bibitem{} King I.R., 1962, AJ 67, 471

\bibitem{} Kofman L., Klypin A., Pogosyan D., Henry P., 1996, ApJ 470,
102

\bibitem{} Lilly S.J., Le Fevre O., Crampton D., Hammer F., Tresse L.,
1995, ApJ 455, 50

\bibitem{} Loveday J., Peterson B.A., Efstathiou G., Maddox S.J., 1992, ApJ
390, 338

\bibitem{} Lubin L.M., Bahcall N.A., 1994, ApJ 426, 513

\bibitem{} Lubin L.M., Postman M., 1996, AJ 111, 1795

\bibitem{} Lucey J.R., Dickens R.J., Mitchell R.J., Dawe J.A., 1983,
MNRAS 203, 545

\bibitem{} Mazure A., Katgert P., den Hartog R., et al., 1996, A\&A
310, 31 : paper II

\bibitem{} Merrifield M.R., Kent S.M., 1989, AJ 98, 351

\bibitem{} Meyer S.L., 1975, ''Data Analysis for Scientists and
Engineers'', p.352, John Wiley \& Sons, New York

\bibitem{} Michie R.W., Bodenheimer P., 1963, MNRAS 126, 269

\bibitem{} Navarro J.F., Frenk C.S., White S.D.M., 1995, MNRAS 275,
720

\bibitem{} Navarro J.F., Frenk C.S., White S.D.M., 1996, MNRAS 462,
563

\bibitem{} Navarro J.F., Frenk C.S., White S.D.M., 1997, ApJ 490, 493

\bibitem{} Nelder J.A., Mead R., 1965, Comput. J. 13, 317

\bibitem{} Perea J., Del Olmo M., Moles M., 1990, A\&A 237, 319

\bibitem{} Plionis M., Barrow J.D., Frenk C.S., 1991, MNRAS 249, 662

\bibitem{} Sarazin C.L., 1980, ApJ 236, 75

\bibitem{} Silverman B., 1986, Density Estimation for Statistics and
Data Analysis, London: Chapman and Hall

\bibitem{} Teague P.F., Carter D., Gray P.M., 1990, ApJS 72, 715

\bibitem{} van Albada T.S., 1982, MNRAS 201, 939

\bibitem{} van Kampen E., 1995, MNRAS 273, 295

\bibitem{} Walter C., Klypin A., 1996, ApJ 462, 13

\end{thebibliography}
\end{document}